\pdfoutput=1

\documentclass[11pt]{article}

\usepackage[final]{acl}

\usepackage{times}
\usepackage{latexsym}

\usepackage[T1]{fontenc}

\usepackage[utf8]{inputenc}

\usepackage{microtype}

\usepackage{inconsolata}

\usepackage{graphicx}

\usepackage{amssymb}

\usepackage{listings}
\usepackage{multirow}
\usepackage{array}

\usepackage{amsmath}

\usepackage{makecell}

\title{A Benchmark for Audio Reasoning Capabilities of Multimodal Large Language Models}

\author{
  \textbf{Iwona Christop\textsuperscript{1}},
  \textbf{Mateusz Czy{\.z}nikiewicz\textsuperscript{2}},
  \textbf{Pawe{\l} Sk{\'o}rzewski\textsuperscript{1}},
  \textbf{{\L}ukasz Bondaruk\textsuperscript{2}},
\\
  \textbf{Jakub Kubiak\textsuperscript{2}},
  \textbf{Marcin Lewandowski\textsuperscript{2}},
  \textbf{Marek Kubis\textsuperscript{1}}
\\
\\
  \textsuperscript{1}Adam Mickiewicz University, ul. Uniwersytetu Pozna{\'n}skiego 4, 61-614 Pozna{\'n}, Poland
  \\
  \textsuperscript{2}Samsung R\&D Institute Poland, Plac Europejski 1, 00-844 Warszawa, Poland
}

\begin{document}

\maketitle

\begin{abstract}
The present benchmarks for testing the audio modality of multimodal large language models concentrate on testing various audio tasks such as speaker diarization or gender identification in isolation. Whether a multimodal model can answer the questions that require reasoning skills to combine audio tasks of different categories, cannot be verified with their use. To address this issue, we propose Audio Reasoning Tasks (ART), a new benchmark for assessing the ability of multimodal models to solve problems that require reasoning over audio signal.
\end{abstract}

\section{Introduction}

Multimodal large language models (MLLMs) expand upon conventional language models by providing
capabilities to process visual and audio data. While there exist elaborate benchmarks that require
advanced reasoning skills to solve problems grounded in vision~\citep{lee2024vhelm}, the prevalent
paradigm for evaluating auditory modality of MLLMs relies on testing selected capabilities in
isolation. Unfortunately, this approach does not guarantee that problems that require reasoning
across different categories of audio tasks are solved with satisfactory performance, even if
superhuman performance of the model is reported for separate tasks such as speech recognition,
acoustic scene classification or audio captioning. This issue is of particular concern considering
that a common practice for building MLLMs involves using textual and audio components pre-trained
separately to develop the final model~\citep[e.g.,][]{llama3, ultravox06}.

To address deficiencies of current approaches proposed for evaluating auditory modality of MLLMs,
we introduce Audio Reasoning Tasks~(ART), a new benchmark that comprises a set of tasks designed to assess MMLMs' ability to solve problems that require combining diverse skills in understanding audio signals with the ability to reason over their combination. The benchmark is designed to be comprehensible by a person without hearing impairment. The evaluation results reported in Section~\ref{sec:experiments} show that although the proposed tasks can be easily solved by a human, they present a challenge for multimodal large language models.

\section{Related Work}

Benchmarks aimed at evaluating the audio capabilities of language models have been developed for several years. However, early approaches, such as ASR-GLUE~\citep{feng22asr}, had limited capacity for assessing reasoning across different audio tasks. ASR-GLUE adapts GLUE~\citep{wang-etal-2018-glue}, a benchmark for evaluating natural language understanding models, by converting its textual input into speech. Specifically, five NLU tasks from GLUE were selected, and their textual input was converted to audio by recording human speakers under various acoustic conditions. While valuable for evaluating NLU model robustness to ASR errors, this approach does not assess the audio reasoning capabilities of the models. Although ASR-GLUE provides a valuable starting point for audio modality evaluation, it should be noted that textual LLMs are evaluated using more comprehensive benchmarks that include complex reasoning tasks~\citep{harness,liang2023holistic}. Additionally, ASR-GLUE focuses solely on natural language understanding, omitting other audio-oriented tasks.

\citet{wang-etal-2025-audiobench} proposed AudioBench, a benchmark designed for evaluating Audio LLMs. It includes 26 datasets, both new and existing, and focuses on three main areas: speech understanding, audio scene understanding, and audio classification. However, these categories are evaluated in isolation, which means that it cannot be a comprehensive evaluation of audio reasoning. The benchmark uses a system formed from Whisper~\citep{whisper} and Llama-3~\citep{llama3} as a baseline. With the exception of automatic speech recognition tasks, evaluated with word error rate, the evaluation process primarily relies on the LLM-as-a-judge paradigm~\citep{zheng23llmaaj} for performance assessment, raising potential bias concerns. Although AudioBench aims to evaluate key aspects of general-purpose Audio LLMs, it mainly focuses on the accuracy of individual audio tasks.

AIR-Bench, introduced by~\citet{yang-etal-2024-air}, evaluates Audio LLMs in two dimensions: \emph{foundation} and \emph{chat}. The \emph{foundation} component consists of 19 single-choice question tasks related to speech, sound, and music. The \emph{chat} component features open-ended questions covering the same categories. The authors also create a category of mixed problems by combining speech-based tasks with sound and music tasks. The benchmark exclusively uses a closed-source LLM (GPT-4,~\citealp{gpt4}) as a judge in the evaluation process, without considering alternative evaluation methods. A pipeline consisting of GPT-4 combined with Whisper ASR~\citep{whisper} also serves as the baseline system. The study does not address the potential impact of using the same LLM as a judge and as an evaluated model, a known issue in the literature (\citealp{liu-etal-2023-g}, \citealp{liu24narcissistic}). Similar to AudioBench, AIR-Bench focuses on evaluating the accuracy of individual tasks rather than reasoning about the entire audio input, therefore it cannot be classified as a proper audio-reasoning benchmark.

SALMon, presented by~\citet{salmon}, concentrates on features such as background noise, emotion, speaker identity, and room impulse response. It evaluates both intra-recording consistency and alignment with spoken text. The task involves comparing the model's likelihood assignments to two samples, one of which is more plausible. For consistency evaluation, one sample includes a feature shift mid-recording (e.g. background noise or speaker voice). The two samples being compared can differ in terms of background sound, e.g., the same conversation about opening a bank account might be accompanied by street noise in one instance, while in another, it could feature a more fitting background sound, such as that of a bank environment. While humans easily identify the more coherent sample, models often struggle, highlighting an interesting, previously underexplored area. However, unlike our work, the SALMon benchmark focuses on evaluating a single aspect of the model rather than comprehensively assessing whether the model can perform inference based on the source recording.

Audio modality performance results released by commercial providers of
multimodal LLMs are rather limited. GPT-4o reports only automatic speech recognition and audio translation performance\footnote{\url{https://openai.com/index/hello-gpt-4o}} using WER and BLEU metrics, with CoVoST-2~\citep{wang2020covost2} as the evaluation dataset. Gemini 1.5~\citep{geminiteam2024gemini} follows a similar approach, using internal datasets alongside MLS~\citep{mls}, FLEURS~\citep{fleurs}, and CoVoST-2. Llama 3.1~\citep{llama3}, a prominent open-weight multimodal model, also reports speech recognition and audio translation results on MLS, LibriSpeech~\citep{librispeech}, VoxPopuli~\citep{voxpopuli}, FLEURS, and CoVoST-2.

A common characteristic of the audio modality benchmarks discussed above is their emphasis on tasks that assess fundamental audio processing abilities, such as audio classification or sound event detection, in isolation. However, they tend to lack comprehensive, end-to-end evaluation of audio-based inference and reasoning, particularly within a cross-modal context, which is still an underexplored area. This situation has started to change only recently. An example could be the newly introduced MMAU benchmark~\citep{sakshi2024mmaumassivemultitaskaudio}, designed to assess multimodal audio understanding models on tasks requiring expert knowledge and complex reasoning. It comprises 10\,000 audio clips paired with human-annotated questions and answers related to speech, environmental sounds, and music. The benchmark questions focus on information extraction and reasoning. While MMAU covers 27 different skills, none of the tasks proposed in our benchmark has a counterpart in MMAU.\footnote{A comparison of ART and MMAU tasks is provided in Appendix~\ref{app:comparison-art-mmau}.} Furthermore, compared to the benchmark presented in this paper, MMAU has several limitations. Specifically, a number of tasks rely on specialized musical knowledge, such as harmony, chord progression, or melody structure, which can hinder comprehensive error analysis by the broader expert community. A more significant issue is the separation of modalities: the questions about the recordings are in text form and are not an integral part of the audio input. In contrast, in our benchmark, the questions were converted to speech and integrated seamlessly with the input audio. Furthermore, in MMAU, questions are presented as closed-choice tests, with additional answers (distractors) generated using a closed-source language model (GPT-4). Taking into consideration that LLMs tend to recognize their own output~\cite[cf.][]{panickssery2024favor}, this can affect the evaluation process in an unpredictable way. By comparison, \emph{Yes/No} variant of our benchmark eliminates the risk of potential bias in evaluating GPT-4-based models due to the construction of the answer set.

\section{Tasks}
\label{sec:tasks}

\begin{figure}[ht]
  \centering
  \includegraphics[width=0.8\columnwidth]{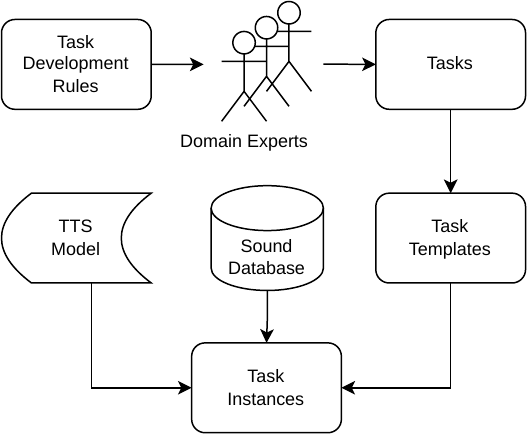}
  \caption{Task preparation process.}
  \label{fig:task-preparation}
\end{figure}

The preparation of the dataset is a multi-stage procedure. First, we formulate a set of rules for the tasks to be included in the dataset. We then survey a group of domain experts to recommend candidate tasks that align with these rules. Next, we eliminate any candidate tasks that cannot be reliably evaluated. Finally, we operationalize the proposed tasks by developing a set of templates that are instantiated with sound samples and synthesized speech in the last stage. The whole process is outlined in Figure~\ref{fig:task-preparation}.

\subsection{Task Selection}

As the performance of MLLMs in tasks that address specific audio problems such as speaker diarization or gender identification can already be tested with the use of targeted datasets and the benchmarks that aim to evaluate AudioLLMs incorporate task-specific datasets~\cite[e.g.][]{wang-etal-2025-audiobench,yang-etal-2024-air}, we refrained from testing the qualities of MLLMs in isolation. Instead, we focused on tasks that combine different sound phenomena to assess the capability of MLLMs to reason over diverse input signals. Thus, to be included in the ART dataset, the task has to obey the following rule:

\begin{quote}
  \textbf{Rule 1:} The task should not be solvable by an LLM that consumes the output of a single specialized module that approaches a specific task and ignores all other sound phenomena presented in the audio signal.
\end{quote}
This rule excludes tasks that can be solved by an LLM acting on speech transcription. Furthermore, it eliminates tasks that can be solved using an audio captioning model followed by a question answering system, effectively excluding the majority of \emph{Chat} tasks proposed by~\citet{yang-etal-2024-air}.

To simplify error analysis for independent researchers who decide to use our benchmark, we also assumed that the proposed tasks should not rely on the superhuman performance of the model or the competence of highly skilled individuals, such as musicians or sound engineers. Thus, the second rule that the tasks have to obey is:

\begin{quote}
  \textbf{Rule 2:} The task should be approachable by a person without professional training.
\end{quote}

To gather tasks that obey the aforementioned rules, we surveyed a group of experienced speech and natural language processing engineers that included the authors of this paper. Initially, we collected 25 candidate tasks from 7 participants in total. In our effort to create a benchmark that produces results easily verifiable by an unskilled person, we decided to exclude from our candidate set all tasks that could lead to different outcomes due to individual variability. Thus, we rejected tasks that involved emotion classification or subjective assessment of sound quality. Furthermore, since we intended to provide an option to evaluate the model without depending on another LLM to serve as a judge, we eliminated tasks that cannot be framed as \emph{Yes/No} questions. As a result, our benchmark is composed of nine tasks presented in Table~\ref{tab:task-templates}.

\begin{table*}[t]
  \caption{Descriptions and examples of questions for each task included in the ART benchmark.}
  \label{tab:task-templates}
  \centering
  \begin{tabular}{p{0.24\textwidth}p{0.35\textwidth}p{0.33\textwidth}}
    \hline
    \textbf{Task name} & \textbf{Description} & \textbf{Example} \\
    \hline
    \textbf{Audio Arithmetics} & Performing simple arithmetic reasoning with regard to the sounds
      heard. & Are there as many bell rings as there are cat meows? \\ \hline
    \textbf{Audio Transformation\newline Detection} & Recognizing whether one recording is a
      transformed version of the other. & Is the first recording a sped up version of the second
      recording? \\ \hline
    \textbf{Cross-Recording\newline Language Identification} & Comparison of the languages spoken in
      the recordings. & Is Budapest the capital of the country this speaker comes from? \\ \hline
    \textbf{Cross-Recording\newline Speaker Identification} & Comparison of the speakers in the recordings.
      & Is the same person heard speaking on both recordings? \\ \hline
    \textbf{Selective Text Inference} & Inference based on some uttered content which is selected on
      the basis of the properties of some of the speakers. & Is green the answer to the question
      asked by a man? \\ \hline
    \textbf{Sound Reasoning} & Reasoning based on the recognized sound. & Is the animal that makes
      the following sound bigger than a horse? \\ \hline
    \textbf{Speech Features\newline Comparison} & Comparison of two recordings regarding speech
      features present in them. & Is the second recording the same text but read with a Scottish
      accent? \\ \hline
    \textbf{Text and Sound\newline Reasoning} & Questions that require both sound features and text
      understanding to be answered. & Is the person talking about the following sound? \\ \hline
    \textbf{Text and Temporal\newline Localization Reasoning} & Questions that require both noises
      from localization (surroundings) and text understanding to be answered. & Does the speaker
      describe the acoustic scene that they are in? \\
    \hline
  \end{tabular}
\end{table*}

\subsection{Task Templates}

We relied on the use of templates to generate the tasks. The rationale for adopting this approach was twofold. First, we wanted to control the size of the benchmark. Templates allowed us to easily increase the diversity of questions by expanding sets of possible slot values while maintaining the preferred size of the dataset. Second, as with any dataset released to the public, there is an ongoing risk of training data contamination for models that will be released in the future. Having a set of templates that can be populated with fresh sound samples from undisclosed sources will allow developers of future models to re-run the evaluation procedure while mitigating the risk of obtaining over-optimistic results due to data contamination.
For each task, we developed a set of question templates containing empty slots filled with appropriate values.
The values could be single words, whole sentences, or special tags, later to be replaced with
sounds. During template creation, values were randomly chosen from previously prepared sets in a
manner that allowed us to automatically generate a target answer to the question. An example of a
template is shown in Figure~\ref{fig:template}.\footnote{Task templates are described in detail in Appendix~\ref{app:task-templates}.}

\begin{figure}[ht]
    \centering
    \includegraphics[width=0.95\columnwidth]{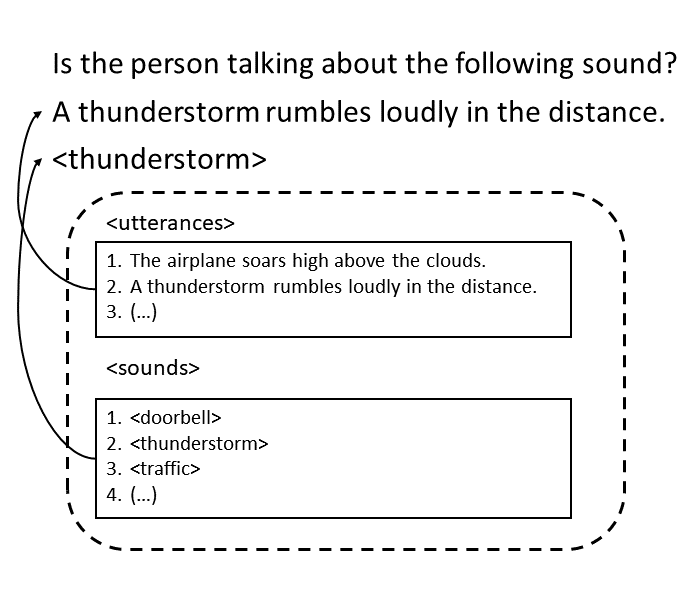}
    \caption{Example of a template. Both the sentence and the sound are chosen randomly from predefined
      lists; the target answer can be inferred from these values. Proper speakers for voice cloning
      for each part of the template are selected.}
    \label{fig:template}
\end{figure}

\begin{table*}[htb]
  \caption{Summary statistics of the ART benchmark.}
  \label{tab:task-instances}
  \centering
  \begin{tabular}{p{0.1\textwidth}cccccc}
    \hline
    \textbf{Task} & \textbf{\# Samples} & \textbf{\# Templates} & \textbf{\# Speakers} &
    \textbf{\# Utterances} & \textbf{\# Sounds} & \textbf{Total length} \\
    \hline
    AA & 1000 & 6 & N/A & N/A & 5 & 3h 46m 10s \\
    ATD & 1000 & 4 & N/A & N/A & 4 & 3h 53m 30s \\
    CRLI & 1000 & 6 & 12 & 12 & N/A & 3h 32m 1s \\
    CRSI & 1000 & 4 & 4 & 8 & N/A & 3h 46m 17s \\
    STI & 1000 & 4 & 4 & 36 & N/A & 3h 9m 47s \\
    SR & 1000 & 15 & N/A & N/A & 20 & 3h 8m 15s \\
    SFC & 1000 & 4 & 10 & 3 & N/A & 2h 37m 7s \\
    TSR & 1000 & 8 & 4 & 16 & 17 & 3h 25m 24s \\
    TTLR & 1000 & 4 & 4 & 15 & 8 & 3h 47s \\
    \hline
    \textbf{Total} & 9000 & 55 & 22 & 86 & 25 & 30h 19m 18s \\
    \hline
  \end{tabular}
\end{table*}

\subsection{Task Instances}

Based on this information, speech samples for voice-cloning were randomly chosen from previously prepared sets with required characteristics. This approach also helped to enhance the diversity of the dataset.

To generate instances of each task, we needed three types of audio recordings -- questions, utterances and sounds.

To unify audio prompts, we synthesized all questions using the same sample for voice cloning. As the goal was to generate intelligible speech, the choice was limited to these samples from the LJ Speech dataset~\citep{ljspeech}, for which the Whisper medium model~\citep{whisper} obtained WER of 0. The chosen samples were used to generate synthetic speech, which was again transcribed to find samples that yielded a WER of 0. The remaining samples were evaluated by human reviewers, and the most natural-sounding sample was selected as the prompt for question synthesis.

Five of the tasks required additional utterances spoken by different speakers. The samples for voice cloning were selected from the GLOBE dataset~\citep{globe}. As with selecting the prompt for the questions, the samples were transcribed using the Whisper medium model, used as prompts for voice cloning, and transcribed again. Thus, after human evaluation, 10 samples (one per speaker) were selected to synthesize utterances.

One of the tasks, Cross-Recording Language Identification, required utterances in languages other than English. For this purpose, we used the VoxPopuli dataset~\citep{voxpopuli}. We selected four languages -- Estonian, Finnish, Hungarian and Polish. For each language, we identified speakers for which WER obtained with the Whisper medium model was equal to 0. We then narrowed our selection to recordings with a duration between 2 and 5 seconds. This process resulted in the selection of three speakers (one utterance each) for each language.

The last type of audio recordings, sounds, included short sounds of e.g. an animal, tunes, and background noises. All sounds were manually selected from samples available under the Creative Commons 0 license in the Freesound dataset~\citep{freesound}, resulting in 13 short sounds, four short tunes, and eight background sounds. The short sounds were manually trimmed as necessary to include a single sound event per sample.

For speech synthesis, we utilized a pipeline based on Voicebox~\citep{voicebox}, a text-to-speech model that demonstrates zero-shot capabilities in reconstructing audio segments from textual inputs and speech prompts. The architecture adapts the transformer model~\citep{transformer}, with modifications including the use of rotary positional embedding~\citep{roformer} instead of ALiBi self-attention bias~\citep{alibi}. To predict token durations, we employ a DurationPredictor model, similar to Voicebox but smaller in size. We also utilize CTC-based forced alignment to discover token durations in an unsupervised manner. A HiFi-GAN vocoder~\citep{hifigan} is used to map audio features to speech, consisting of a fully convolutional generator and two discriminators. The input tokens are represented as phonetic labels, and mel-scale spectrograms with 80 channels are used as audio features.

\begin{table*}[t]
  \caption{Results of model evaluation on the ART benchmark using \emph{Yes/No} approach.}
  \label{tab:results-yes-no}
  \centering
  \begin{tabular}{p{0.33\textwidth}ccc}
    \hline
    \textbf{Model} & \textbf{\% Relevant} & \textbf{Absolute Accuracy} & \textbf{Relative Accuracy} \\
    \hline
    Whisper + Llama & 99.91 & 0.5404 & 0.5408 \\
    Whisper + Qwen & \textbf{99.94} & \textbf{0.5621} & \textbf{0.5625} \\
    \hline
    Audio Flamingo 3 & \textbf{100.00} & \textbf{0.5473} & \textbf{0.5473} \\
    GAMA & 42.56 & 0.2155 & 0.5064 \\
    Qwen-Audio-Chat & 64.09 & 0.3312 & 0.5168 \\
    Qwen2-Audio \scriptsize{(zero-shot)} & 85.41 & 0.4431 & 0.5188 \\
    Qwen2-Audio \scriptsize{(one-shot, same template)} & 87.30 & 0.4710 & 0.5395 \\
    Qwen2-Audio \scriptsize{(one-shot, different template)} & 68.08 & 0.3526 & 0.5179 \\
    Ultravox v0.4.1 & 89.19 & 0.4682 & 0.5250 \\
    Ultravox v0.6 & 99.74 & 0.5309 & 0.5323 \\
    \hline
  \end{tabular}
\end{table*}

After obtaining all necessary samples of questions, utterances, and sounds, the audio recordings were automatically merged according to the prepared template configurations. All recordings were normalized to~-20~dBFS. The duration of silence between the question and the utterance or sound was manually adjusted to ensure that audio prompts sound natural and that the gaps between recordings are sufficient to distinguish them. The short sounds and tunes concatenated with a question were truncated to the maximum duration of~5~seconds, and a fade-out was applied to eliminate sudden volume changes. In the case of instances that required an overlay of background sounds, the underlying audio was attenuated by~20~dB to ensure speech intelligibility, and both fade-in and fade-out were applied.

The consolidation of templates resulted in~9\,000~samples that constitute the final dataset, which amounts to over~30~hours of audio, as shown in Table~\ref{tab:task-instances}.
The prepared dataset is fully balanced across labels and tasks. Of the~9\,000 total samples, 4\,500 have the expected answer of \emph{Yes} and~4\,500 have the expected answer of \emph{No}. This balance is maintained within each of the nine tasks, with~1\,000 samples evenly split between~500 \emph{Yes} and~500 \emph{No} instances. Furthermore, the balance was maintained at the level of task templates wherever possible. For instance, in the \textbf{Audio Transformation Detection} task, template \textbf{ATD\_0} has 56 \emph{Yes} and 56 \emph{No} samples; template \textbf{ATD\_1} has 56/56; template \textbf{ATD\_2} has 112/112; and template \textbf{ATD\_3} has 276/276. This ensures a uniform distribution of labels across tasks.

\section{Experiments}
\label{sec:experiments}

\begin{table*}[htbp]
  \caption{Absolute accuracy per task on the ART benchmark using \emph{Yes/No} approach.}
  \label{tab:yn-results-per-task}
  \centering
  \begin{tabular}{m{0.2\textwidth}ccccccccc}
    \hline
    \textbf{Model} & \textbf{AA} & \textbf{ATD} & \textbf{CRLI} & \textbf{CRSI} & \textbf{STI} &\textbf{SR} & \textbf{SFC} & \textbf{TSR} & \textbf{TTLR} \\
    \hline
    Whisper + Llama & 0.505 & 0.483 & 0.458 & \textbf{0.505} & \textbf{0.665} & 0.525 & \textbf{0.557} & 0.643 & 0.521 \\
    Whisper + Qwen & \textbf{0.510} & \textbf{0.504} & \textbf{0.551} & 0.501 & 0.625 & \textbf{0.654} & 0.532 & \textbf{0.653} & \textbf{0.528} \\
    \hline
    Audio Flamingo 3 & \textbf{0.516} & 0.492 & 0.569 & \textbf{0.517} & 0.554 & \textbf{0.700} & 0.494 & 0.566 & \textbf{0.517} \\
    GAMA & 0.036 & 0.376 & 0.276 & 0.349 & 0.338 & 0.050 & 0.299 & 0.068 & 0.147 \\
    Qwen-Audio-Chat & 0.498 & \textbf{0.521} & 0.008 & 0.123 & 0.201 & 0.549 & 0.181 & 0.525 & 0.375 \\
    Qwen2-Audio\newline \scriptsize{(zero-shot)} & 0.488 & 0.517 & 0.501 & 0.212 & 0.532 & 0.432 & 0.524 & 0.358 & 0.425 \\
    Qwen2-Audio\newline \scriptsize{(one-shot, same template)} & 0.493 & 0.282 & 0.517 & 0.454 & 0.534 & 0.517 & \textbf{0.527} & 0.543 & 0.373 \\
    Qwen2-Audio\newline \scriptsize{(one-shot, different template)} & 0.441 & 0.245 & 0.464 & 0.340 & 0.477 & 0.176 & 0.467 & 0.392 & 0.172 \\
    Ultravox v0.4.1 & 0.475 & 0.288 & 0.516 & 0.489 & 0.478 & 0.438 & 0.507 & 0.511 & 0.512 \\
    Ultravox v0.6 & 0.480 & 0.501 & \textbf{0.579} & 0.504 & \textbf{0.559} & 0.526 & 0.521 & \textbf{0.590} & 0.513 \\
    \hline
    \textbf{Average} & 0.436 & 0.413 & 0.430 & 0.386 & \textbf{0.490} & 0.430 & 0.457 & 0.476 & 0.396 \\
    \hline
  \end{tabular}
\end{table*}

\subsection{Setup}
\label{sec:experiments-setup}

We conducted a series of experiments evaluating multimodal models on prepared tasks. Two different approaches were adopted:
\begin{itemize}
  \item \emph{Yes/No} -- where the model was instructed to answer only \emph{Yes} or \emph{No},
  \item \emph{Descriptive} -- where the form of the answer was not specified and the model was allowed to give a descriptive answer.
\end{itemize}
As all the tasks were designed to have either an affirmative or negative answer, we focused on \emph{Yes/No} approach in this section. However, to gain better understanding of the reasons why the models fail to accomplish the tasks, we also studied the open-ended responses yielded by \emph{Descriptive} approach with summary scores reported in Table~\ref{tab:results-descriptive} and the detailed analysis given in Appendix~\ref{app:descriptive-results}.

To ensure the reliability of the results, the responses obtained using \emph{Yes/No} approach were automatically evaluated. If the answer was \emph{Yes} or \emph{No}, it was marked as relevant, and irrelevant otherwise. The responses marked as relevant were further classified as correct or incorrect. As LLM-as-a-judge is a commonly used evaluation method, we decided to use it in the \emph{Descriptive} approach. For this purpose, we used two models -- Llama-3.3-70B-Instruct~\citep{llama3} and Qwen3-32B~\citep{yang2025qwen3technicalreport}. In the prompt, the model was instructed that it would receive a question, an expected answer and the received answer. Its task was to determine whether the generated answer was relevant or not, and to state whether it was correct or why it was labeled as irrelevant. The full prompt used for this evaluation is shown in Appendix~\ref{app:experimental-setup}, along with the inference parameters.

\subsection{\emph{Yes/No} approach}
\label{sec:experiments-yes-no}

The results of evaluation using \emph{Yes/No} approach are shown in Table~\ref{tab:results-yes-no}. The relative accuracy is defined as the accuracy calculated only on relevant responses. In this case, the models were instructed to answer only \emph{Yes} or \emph{No}. The inference on each of the models was run five times and the results were averaged to assess whether the models exhibit superiority over random guessing.

Two cascaded systems where evaluated, both using Whisper Large v3~\citep{whisper} to obtain transcriptions of the audio prompts. First system used Llama-3.3-70B-Instruct~\citep{llama3} to answer the questions, and the second one utilized Qwen3-32B~\citep{yang2025qwen3technicalreport}. Both cascaded systems generated almost~100\% relevant answers, achieving~54.04\% and~56.21\% absolute accuracy, respectively.

Considering MLLMs, Audio Flamingo 3~\citep{goel2025audioflamingo3advancing} is the only model that returned~100\% relevant answers. It also outperformed the other models with the accuracy of~54.73\%. Ultravox-v0.6-Llama3.3-70B~\citep{ultravox06} obtained the second highest number of relevant responses. It managed to achieve absolute accuracy of~53.09\%.
The previous version of this model, Ultravox-v0.4.1-Llama3.1-8B, generated~10\% less relevant
responses, achieving less than~50\% absolute accuracy in all five runs. Qwen2-Audio-7B-Instruct~\citep{chu2023qwenaudioadvancinguniversalaudio} generated only~85.41\% relevant responses in the zero-shot approach. As the authors of this model suggest using the one-shot approach, two additional experiments were performed. In the first experiment, the model was given a sample from the same template as an example. This resulted in less than~2\% more relevant answers and less than~3\% greater accuracy. In the second experiment, the model was given as an example a sample from the same task but a different template. This approach resulted in a~17\% decrease in the number of relevant responses and a~9\% decrease in accuracy compared to zero-shot approach. The lowest results in the Qwen family of models were achieved by Qwen-Audio-Chat~\citep{chu2023qwenaudioadvancinguniversalaudio}. It generated only~64.09\% relevant responses, achieving average absolute accuracy of~33.12\%. The GAMA model~\citep{ghosh-etal-2024-gama} reached only~21.55\% accuracy, which was the lowest in the evaluation using \emph{Yes/No} approach.

The results per task achieved by the models for the \emph{Yes/No} approach are shown in Table~\ref{tab:yn-results-per-task}. Based on these, the \textbf{Cross Recording Speaker Identification} proved to be the hardest task -- the average absolute accuracy is~38.63\% and none of the models achieved accuracy higher than~50\% in any of the five runs. Of these, the best average absolute accuracy was achieved for the \textbf{Selective Text Inference} task.

\begin{table*}[t]
  \caption{Results of the model evaluation on the ART benchmark using \emph{Descriptive} approach.}
  \label{tab:results-descriptive}
  \centering
  \begin{tabular}{p{0.22\textwidth}|cc|cc|c}
    \hline
    & \multicolumn{2}{c|}{\textbf{Llama 3.3}} & \multicolumn{2}{c|}{\textbf{Qwen3}} & \\
    \textbf{Model} & \textbf{Relevant} & \textbf{Accuracy} & \textbf{Relevant} & \textbf{Accuracy}
      & \textbf{Agreement}\\
    \hline
    Whisper + Llama & 62.62\% & 0.3535 & 62.17\% & 0.4834 & 74.07\% \\
    Whisper + Qwen & \textbf{89.68\%} & \textbf{0.4882} & \textbf{87.97\%} & \textbf{0.5697}
      & \textbf{76.09\%} \\
    \hline
    Audio Flamingo 3 & 74.07\% & 0.3536 & 76.99\% & 0.3524 & 82.10\% \\
    GAMA & 1.31\% & 0.0089 & 2.40\% & 0.0154 & \textbf{97.30\%} \\
    Qwen-Audio-Chat & 14.50\% & 0.0778 & 19.89\% & 0.0548 & 86.04\% \\
    Qwen2-Audio\newline \scriptsize{(zero-shot)} & 92.26\% & 0.4490 & 94.94\% & 0.4683 & 82.43\% \\
    Qwen2-Audio\newline \scriptsize{(one-shot, same template)} & \textbf{93.94\%} & \textbf{0.4735}
      & \textbf{96.36\%} & \textbf{0.4996} & 83.31\% \\
    Qwen2-Audio\newline \scriptsize{(one-shot, different template)} & 93.16\% & 0.4495 & 95.28\%
      & 0.4834 & 81.30\% \\
    Ultravox v0.4.1 & 60.20\% & 0.2398 & 50.59\% & 0.1692 & 76.19\% \\
    Ultravox v0.6 & 79.42\% & 0.3952 & 77.69\% & 0.4425 & 77.14\% \\
    \hline
  \end{tabular}
\end{table*}

\subsection{\emph{Descriptive} approach}
\label{sec:experiments-descriptive}

Table~\ref{tab:results-descriptive} shows the results of the experiments performed with the \emph{Descriptive} approach and both Llama and Qwen3 as a judge.

According to both judges, none of the models achieved satisfactory results. Only the cascaded system using Qwen3 achieved absolute accuracy higher than~50\%, but only when it evaluated itself. The agreement between judges exceeded~74\% in all cases, and the highest value reached was~97.3\%. However, this case involved the GAMA model, which had an accuracy of less than~2.5\%.

Among the MLLMs, Qwen2-Audio in the one-shot approach using the same template as an example, achieved the best results -- Llama 3.3 and Qwen3 judged it achieved~47.35\% and~49.96\% accuracy, respectively. This experiment also resulted in the highest fraction of relevant answers.

It is worth noting that Llama 3.3 recognized more relevant answers when judging itself, Qwen3, and both Ultravox v0.4.1 and Ultravox v0.6. On the other hand, Qwen3 was significantly more indulgent when evaluating the MLLMs from the Qwen family of models.

An in-depth analysis of the results per task in the \emph{Descriptive} approach is available in
Appendix~\ref{app:descriptive-results}. In case of cascaded systems, Qwen3 assessed its own performance significantly better across all tasks except for \textbf{Speech Features Comparison}. The best average accuracy was obtained on the \textbf{Speech Features Comparison} task when judged by Llama 3.3, and on \textbf{Text and Sound Reasoning} judged by Qwen3. However, the results on the remaining tasks are significantly understated due to the accuracy obtained by Qwen-Audio-Chat~(5.48-7.78\%) and GAMA~(0.89-1.54\%).

\subsection{Error analysis}

Error analysis for both approaches revealed distinct yet overlapping failure patterns based on human evaluation. Errors in the \emph{Yes/No} approach were primarily driven by failures in audio understanding. Specifically, models did not recognize the presence of speech or sound. Additionally, there was systematic task confusion, which led to transcription or speaker recognition instead of question answering.

In contrast, the \emph{Descriptive} approach exhibited a broader and more heterogeneous set of errors. In addition to frequently failing to recognize the question, the models often produced random, speculative, or language-inconsistent responses and showed stronger biases toward transcription and speaker identification. These behaviors suggest less stable response control in case of the \emph{Descriptive} approach.

Overall, the results suggest that both approaches are susceptible to task misinterpretation. However, in the \emph{Yes/No} approach errors are concentrated around audio perception and task confusion. In contrast, the \emph{Descriptive} approach results in more diverse and less predictable failure behaviors.
A quantitative analysis is provided in Appendix~\ref{app:error-analysis}.

\section{Benchmark Validation}

Although the task collection process described in Section~\ref{sec:tasks} relies on human expertise and clearly defined rules, the task instantiation procedure depends on templates and synthesized speech which may potentially result in a benchmark that is either too complex to be comprehensible by humans or too simple for models to solve.

To address the first issue, we designated ART-H, a subset of the dataset that consists of~24~samples per task resulting in~216~samples in total, that enables manual verification of the evaluation results in under one hour. We presented ART-H prompts to human evaluators, who answered the questions with either \emph{Yes} or \emph{No}. In this way, we achieved a human baseline of~92.90\%, thus confirming the suitability of the prepared tasks. Three tasks -- \textbf{Audio Arithmetics}, \textbf{Selective Text Inference}, and \textbf{Text and Temporal Localization Reasoning} -- turned out to be the easiest. The worst results were obtained on \textbf{Speech Features Comparison} and \textbf{Cross-Recording Language Identification}. The reason may be that these tasks focus on distinguishing accents or languages, which can be challenging for those unfamiliar with a particular dialect. Detailed results of the human evaluation are provided in Appendix~\ref{app:human-evaluation-results}. Furthermore, to verify if the choice of particular samples to be included in ART-H impacts the results in a meaningful way, we evaluated the models under study with respect to randomly sampled 216-element subsets of ART and demonstrated that this procedure results in a standard deviation of less than~$0.035$ in terms of absolute accuracy~(cf. Appendix~\ref{app:sampling}).

The reliance on synthesized samples raises concerns about the impact of the audio quality on the benchmark results. If the benchmark encompasses audio samples of poor quality, the models could underperform due to artefacts in data. Taking into consideration that we use a state-of-the-art TTS model and perform human evaluation of the ART-H subset, this is not the case. On the other hand, high-quality synthesis can potentially lead to overoptimistic results in speech-related tasks. However, the dependence on synthesized speech ensures that any mistakes observed in the models' performance are due to reasoning errors rather than poor or ambiguous input. If a model demonstrates weak performance on clean audio samples, it will likely perform worse in noisy conditions. The results presented in Section~\ref{sec:experiments} show that none of the models attained satisfactory performance with regard to the synthesized data. Therefore, we believe that the use of audio data that exhibit more demanding acoustic conditions can be postponed.

\section{Conclusion}

In this paper, we proposed Audio Reasoning Tasks~(ART), a new benchmark for assessing the performance of MLLMs. Contrary to the existing benchmarks for testing the audio modality of MLLMs that aim at evaluating various audio capabilities in isolation, our dataset encompasses tasks that, to be solved, require combining a diverse set of skills grounded in the audio domain. The benchmark was designed to be solvable by an unskilled person without hearing impairment. The experiments that were conducted showed that while the benchmark is easily comprehensible by humans, it still poses a challenge for the state-of-the-art open-weight models that we investigated.

\section{Limitations}

While our benchmark can be employed to determine if the model under study is capable of reasoning over an audio signal, the proposed set of tasks cannot be considered complete. Therefore, a successful completion of this test cannot guarantee that the model's audio reasoning skills are at the human level.

\section*{Acknowledgments}

This research was partially funded by the \emph{SPEACAIR: SPEech-Aware Conversational AI Research} project, a cooperation between Adam Mickiewicz University and Samsung Electronics Poland.

\bibliography{custom}

\appendix

\clearpage

\section{ART vs. MMAU tasks}
\label{app:comparison-art-mmau}

\subsection{Audio Arithmetics}

The closest task to \textbf{Audio Arithmetics} is \textbf{Temporal Event Reasoning} which includes questions about how many times a given sound occurred. However, there is only one kind of sound present in the audio recording, while in \textbf{Audio Arithmetics} the sounds occur next to each and the model is required to count both of them and compare, or reason about the evenness of the number of occurrences. There are two different similar tasks -- \textbf{Counting}, which requires counting how many speakers are involved in a conversation, and \textbf{Phonological Sequence Decoding}, which involves counting how many times a given word appears in the audio, but none of them is the exact match to \textbf{Audio Arithmetics}.

\subsection{Audio Transformation Detection}

There is no task similar to \textbf{Audio Transformation Detection}. While \textbf{Temporal Event Reasoning} includes question about identifying the shortest or longest sound, it does not involve comparison of two sounds and detection of modification applied on the audio.

\subsection{Cross-Recording Language Identification}

There is no task similar to \textbf{Cross-Recording Language Identification} -- none of them involves recognition of the language spoken. While \textbf{Counting} requires counting speakers involved in a conversation, no reasoning over their identity is included.

\subsection{Cross-Recording Speaker Identification}

While \textbf{Counting} involves identification of the number of speakers, it does not require comparison of their identities. There is no other task in MMAU that would involve determining whether the same speaker is being heard on both recordings or if the same speakers are involved in two different dialogues, as in \textbf{Cross-Recording Speaker Identification}.

\subsection{Selective Text Inference}

The closest tasks to \textbf{Selective Text Inference} are: \textbf{Key highlight Extraction} and \textbf{Conversational Fact Retrieval}, which involve questions about what speaker 1 or 2 said, and \textbf{Event-Based Knowledge Retrieval}, which requires question answering based on the given utterance. However, none of this tasks involves reasoning based on comparison of speaker characteristics. In \textbf{Selective Text Inference}, in addition to question answering, each template includes identification of speaker's gender or recognition of the utterance's subject.

\subsection{Sound Reasoning}

\textbf{Sound-Based Event Recognition} requires identifying of an event based on audio. \textbf{Ambient Sound Interpretation} includes two sounds -- one of them is mentioned in the question, and the second one is to be identified. \textbf{Eco-Acoustic Knowledge} is based on recognition of the environment and the question often suggests the answer, e.g. waves can be heard and the question asks what natural disaster can be inferred that typically results in significant loss of life and property due to subsequent flooding particularly in coastal regions. \textbf{Event-Based Sound Reasoning} asks what caused a given sound, e.g. howling -- which again suggests the answer, as a howl is often associated with wolves. While all of these tasks involve some kind of reasoning over sounds, none of them is the same as \textbf{Sound Reasoning}, which requires not only sound recognition, but also actual reasoning. For example, the task is to recognize two animals based on their sounds and determining whether both of them are mammals.

\subsection{Speech Features Comparison}

There is no task similar to \textbf{Speech Features Comparison} as none of them considers speakers' characteristics such as accent, age or gender. While \textbf{Counting} involves counting how many speakers can be heard, no reasoning over their characteristics is required.

\subsection{Text and Sound Reasoning}

There is no task that could be a match for \textbf{Text and Sound Reasoning}. While some tasks require reasoning over sounds, as described for \textbf{Sound Reasoning} task, none of them involves reasoning based on both sounds and utterances. For example, the \textbf{Text and Sound Reasoning} task requires recognition of two animals based on a sound and an utterance, and determining whether both of them are smaller than an elephant.

\subsection{Text and Temporal Localization Reasoning}

The most similar task is \textbf{Acoustic Scene Reasoning}, which includes identification of localization where a given sound can be heard. While \textbf{Text and Temporal Localization Reasoning} also asks about the time and place of the recording, it involves reasoning over both speech and background sounds and comparison of the acoustic environment of two speakers.

\section{Task Templates}
\label{app:task-templates}

\subsection{Audio Arithmetics}

Let~$S$ be the predefined set of sounds:

\begin{align*}
  S = \{ &\text{bell ring}, \text{cat meow}, \text{dog bark}, \\
    &\text{elephant trumpet}, \text{horse neigh} \}
\end{align*}

Template~\ref{eq:aa0} requires comparing the number of occurrences of sounds~$s_A$ and~$s_B$. To do this, the model must count how many times sound~$s_A$ and sound~$s_B$ occurred, and then determine if the number is equal.

\begin{align}
  &\text{Are there as many } s_A \text{ as there are } s_B \text{?} \notag \\
  &+ \left( n_A \times s_A \right) + \left( n_B \times s_B \right)
  \tag{AA\_0} \label{eq:aa0}
\end{align}

\noindent
where~$s_A, s_B \in S$, $s_A\neq s_B$; $n_A, n_B \in \langle 1, 4 \rangle$. In this template, the answer is \emph{Yes} if~$n_A = n_B$.

Similarly, in~\ref{eq:aa1}, the model must determine whether sound~$s_A$ occurs twice as much as sound~$s_B$.

\begin{align}
  &\text{Are there twice as many } s_A \text{ as there are } s_B \text{?} \notag \\
  &+ \left( n_A \times s_A \right) + \left( n_B \times s_B \right)
  \tag{AA\_1} \label{eq:aa1}
\end{align}

\noindent
where~$s_A, s_B \in S$, $s_A\neq s_B$; $n_A, n_B \in \langle 1, 4 \rangle$. The answer is \emph{Yes} if~$n_A=2n_B$.

Templates~\ref{eq:aa2} and~\ref{eq:aa3} involve determining whether sound~$s_A$ occurred an even or an odd number of times. In both cases, sound~$s_B$ can occur next to~$s_A$, which requires correct sound recognition. In template~\ref{eq:aa2} the answer is \emph{Yes} if~$n_A\mod 2 = 0$, and in~\ref{eq:aa3} if~$n_A\mod 2 \neq 0$.

\begin{align}
  &\text{Does the } s_A \text{ an even number of times?} \notag \\
  &+ \left( n_A \times s_A \right) + \left( n_B \times s_B \right)
  \tag{AA\_2} \label{eq:aa2} \\
  \notag \\
  &\text{Does the } s_A \text{ an odd number of times?} \notag \\
  &+ \left( n_A \times s_A \right) + \left( n_B \times s_B \right)
  \tag{AA\_3} \label{eq:aa3}
\end{align}

\noindent
where~$s_A, s_B \in S$, $s_A\neq s_B$; $n_A \in \langle 1, 4 \rangle$; $n_B \in \langle 0, 4 \rangle$.

Template~\ref{eq:aa4} requires comparison if sound~$s_A$ occurred more times than sound~$s_B$. Then, the answer is \emph{Yes} if~$n_A > n_B$. Similarly, in template~\ref{eq:aa5} the answer is \emph{Yes} if sound~$s_A$ occurred less times than sound~$s_B$, i.e.~$n_A < n_B$.

\begin{align}
  &\text{Are there more } s_A \text{ than } s_B \text{?} + \left( n_A \times s_A \right) \notag \\
  &+ \left( n_B \times s_B \right)
  \tag{AA\_4} \label{eq:aa4} \\
  \notag \\
  &\text{Are there less } s_A \text{ than } s_B \text{?} + \left( n_A \times s_A \right) \notag \\
  &+ \left( n_B \times s_B \right)
  \tag{AA\_5} \label{eq:aa5}
\end{align}

\noindent
where~$s_A, s_B \in S$, $s_A\neq s_B$; $n_A, n_B \in \langle 1, 4 \rangle$.

\subsection{Audio Transformation Detection}

Let~$S$ be the predefined set of sounds and~$T$ the predefined set of transformations:

\begin{align*}
  S = \{ &\text{circus}, \text{drum}, \text{elevator}, \text{piano} \} \\
  T = \{ &\text{higer}, \text{louder}, \text{lower}, \text{quieter}, \text{reversed}, \\
    &\text{slowed down}, \text{sped up}, \text{truncated} \}
\end{align*}

\noindent
then~$t(s)$ means that transformation~$t\in T$ is applied to sound~$s\in S$.

Template~ATD\_0 involves determining whether first sound is a modified version of the second recording.

\begin{align}
  &\text{Is the first recording a } t \text{ version of the second} \notag \\
  &\text{recording?} + t_A(s_A) + s_B
  \tag{ATD\_0.1} \label{eq:atd01}
\end{align}

\noindent
where~$s_A, s_B\in S$; $t, t_A\in T$. The answer is \emph{Yes} if~$s_A = s_B \land t = t_A$.

To include additional affirmative answers, a second version of this template,~\ref{eq:atd02}, was designed. In this case, transformation~$t_B$ is applied to sound~$s_B$. The answer is \emph{Yes} if~$t_B$ is the inverse transformation of~$t$, i.e.~$t_B = t^{-1}$, and~$s_A = s_B$.

\begin{align}
  &\text{Is the first recording a } t \text{ version of the second} \notag \\
  &\text{recording?} + s_A + t_B(s_B)
  \tag{ATD\_0.2} \label{eq:atd02}
\end{align}

\noindent
where~$s_A, s_B\in S$; $t, t_B\in T \setminus \{$reversed, truncated$\}$.

Similarly, two versions of template~ATD\_1 were designed. \ref{eq:atd11} requires determining whether sound~$s_B$ is a transformed version of~$s_A$. The answer is \emph{Yes} if~$s_A = s_B \land t = t_B$.

\begin{align}
  &\text{Is the second recording a } t \text{ first recording?} \notag \\
  &+ s_A + t_B(s_B)
  \tag{ATD\_1.1} \label{eq:atd11}
\end{align}

\noindent
where~$s_A, s_B\in S$; $t, t_B\in T$.

In~\ref{eq:atd12}, transformation~$t_A$ is applied to~$s_A$, and the answer is affirmative if~$s_A = s_B \land t_A = t^{-1}$.

\begin{align}
  &\text{Is the second recording a } t \text{ first recording?} \notag \\
  &+ t_A(s_A) + s_B
  \tag{ATD\_1.2} \label{eq:atd12}
\end{align}

\noindent
where~$s_A, s_B\in S$; $t, t_A\in T \setminus \{$reversed, truncated$\}$.

Template~\ref{eq:atd2} requires determining if one of the recordings is the same as the other one but with transformation~$t$ applied.

\begin{align}
  &\text{Is one of the recordings a } t \text{ version of the other} \notag \\
  &\text{one?} + t_A(s_A) + t_B(s_B)
  \tag{ATD\_2} \label{eq:atd2}
\end{align}

\noindent
where~$s_A, s_B\in S$; $t_A, t_B\in T \cup \{\text{none}\}$; $\neg (t_A = \text{none} \land t_B = \text{none})$. In this case, the answer is \emph{Yes} if~$s_A = s_B \land \left(t \in \{ t_A, t_B, t_A^{-1}, t_B^{-1} \}\right)$.

In case of template~\ref{eq:atd3}, transformations are applied to both sounds, and the model must determine whether the sounds are the same but the transformations differ. Then, the answer is affirmative if~$s_A = s_B \land t_A \neq t_B$.

\begin{align}
  &\text{Are these two recordings the same with} \notag \\
  &\text{different transformations?} + t_A(s_A) \notag \\
  &+ t_B(s_B)
  \tag{ATD\_3} \label{eq:atd3}
\end{align}

\noindent
where~$s_A, s_B\in S$; $t_A, t_B\in T$.

\subsection{Cross-Recording Language Identification}

Let~$L$ be the predefined set of languages and~$S$ the predefined set of speakers:

\begin{align*}
  L &= \{ \text{Estonian}, \text{Finnish}, \text{Hungarian}, \text{Polish} \} \\
  S &= \{ l_i \mid l \in L, \; i \in \{ 1, 2, 3 \} \}
\end{align*}

\noindent
then

\begin{align*}
  \mathrm{language}(l_i) &:= l \text{ for all } l \in L, \; i \in \{1, 2, 3\}
\end{align*}

Template~\ref{eq:crli0} requires determining if all of the three people in the recording speak the same language.

\begin{align}
  &\text{Do all the people in the recording speak} \notag \\
  &\text{the same language?} + s_A + s_B \notag \\
  &+ s_C
  \tag{CRLI\_0} \label{eq:crli0}
\end{align}

\noindent
where~$s_A, s_B, s_C \in S$; $s_A \neq s_B \neq s_C$. The answer is \emph{Yes} if~$\mathrm{language}(s_A) = \mathrm{language}(s_B) = \mathrm{language}(s_C)$.

Template~\ref{eq:crli1} is similar, but consists of only two recordings. In this case, the answer is \emph{Yes} if~$\mathrm{language}(s_A) = \mathrm{language}(s_B)$.

\begin{align}
  &\text{Can two different languages be heard on} \notag \\
  &\text{the recording?} + s_A + s_B
  \tag{CRLI\_1} \label{eq:crli1}
\end{align}

\noindent
where~$s_A, s_B \in S$; $s_A \neq s_B$.

Template~\ref{eq:crli2} is the opposite of~\ref{eq:crli0}, i.e. the model is must assess whether the speakers use three different languages.

\begin{align}
  &\text{Can three different languages be heard on} \notag \\
  &\text{the recording?} + s_A + s_B + s_C
  \tag{CRLI\_2} \label{eq:crli2}
\end{align}

\noindent
where~$s_A, s_B, s_C \in S$ and~$s_A \neq s_B \neq s_C$. For~\ref{eq:crli2}, the answer is \emph{Yes} if~$\mathrm{language}(s_A) \neq \mathrm{language}(s_B) \neq \mathrm{language}(s_C)$.

Template~\ref{eq:crli3} requires recognition of the speaker's country of origin and determining if the given city is it's capital. Let~$C$ be the predefined set of cities:

\begin{align*}
  C &= \{ \text{Budapest}, \text{Helsinki}, \text{Tallinn}, \text{Warsaw} \}
\end{align*}

\noindent
then~$\mathrm{is\_capital}(c, s)$ implies that city~$c\in C$ is the capital city of the speaker~$s\in S$ country of origin.

\begin{align}
  &\text{Is } c \text{ the capital of the country this speaker} \notag \\
  &\text{comes from?} + s
  \tag{CRLI\_3} \label{eq:crli3}
\end{align}

\noindent
where~$c \in C$, $s \in S$. In this case, the answer is affirmative if~$\mathrm{is\_capital}(c, s)$.

Template~\ref{eq:crli4} involves assessing whether both speakers~$s_A$ and~$s_B$ are from the same country. Since each language in set~$L$ is an official language in one country, obtaining the correct answer comes down to comparing the languages spoken by the speakers, i.e. the answer is \emph{Yes} if~$\mathrm{language}(s_A) = \mathrm{language}(s_B)$.

\begin{align}
  &\text{Are both speakers from the same country?} \notag \\
  &+ s_A + s_B
  \tag{CRLI\_4} \label{eq:crli4}
\end{align}

\noindent
where~$s_A, s_B \in S$, $s_A \neq s_B$.

In template~\ref{eq:crli5}, the model must verify whether the given country borders the speaker's  country of origin. Let~$N$ be the predefined set of countries:

\begin{align*}
  N = \{ &\text{Belarus}, \text{Estonia}, \text{Finland}, \text{Latvia}, \\
    &\text{Norway}, \text{Slovakia}, \text{Ukraine} \}
\end{align*}

\noindent
then~$\mathrm{borders}(n, s)$ implies that country~$n \in N$ borders speaker~$s \in S$ country of origin.

\begin{align}
  &\text{Does } n \text{ border the speaker's country of} \notag \\
  &\text{origin?} + s
  \tag{CRLI\_5} \label{eq:crli5}
\end{align}

\noindent
where~$s \in S$, $n \in N$. The answer to~\ref{eq:crli5} is \emph{Yes} if~$\mathrm{borders}(n, s)$.

\subsection{Cross-Recording Speaker Identification}

Let~$S$ be the predefined set of speakers:

\begin{equation*}
  S = \{s_1, s_2, s_3, s_4\}
\end{equation*}

\noindent
and~$U$ the predefined set of utterances:

\begin{itemize}
  \item ``The soft glow of the morning sun filtered through the curtains.''
  \item ``The rain tapped softly against the window.''
  \item ``The birds chirped a gentle melody.''
  \item ``It was a perfect moment to pause and breathe deeply.''
\end{itemize}

\noindent
then~$s(u)$ means that utterance~$u \in U$ is spoken by speaker~$s \in S$.

Templates~\ref{eq:crsi0} and~\ref{eq:crsi1} require determining whether both utterances~$u_A$ and~$u_B$ are spoken by the same person, i.e. the answer is affirmative if~$s_A = s_B$.

\begin{align}
  &\text{Are both recordings spoken by the same} \notag \\
  &\text{person?} + s_A(u_A) + s_B(u_B)
  \tag{CRSI\_0} \label{eq:crsi0} \notag \\
  \notag \\
  &\text{Is the same person heard speaking on both} \notag \\
  &\text{recordings?} + s_A(u_A) + s_B(u_B)
  \tag{CRSI\_1} \label{eq:crsi1}
\end{align}

\noindent
where~$s_A, s_B\in S$; $u_A, u_B\in U$; $u_A\neq u_B$.

Template~\ref{eq:crsi2} involves assessing whether the same people are speaking in both dialogues~$d_A$ and~$d_B$.

Let~$D$ be the predefined set of dialogues:

\begin{itemize}
  \item ``Did you remember to water the plants today?'' \\
    ``Oh no, I completely forgot!''
  \item ``Did you hear about the new caf\'e downtown?'' \\
    ``No, what's it like?''
  \item  ``Are we meeting at 7 tonight?'' \\
    ``I thought it was 7:30?''
  \item ``Did you finish the group project?'' \\
    ``Almost, I just need to finalize the slides.''
\end{itemize}

\noindent
then~$d(s_A, s_B)$ means that the first utterance in dialogue~$d \in D$ is spoken by the speaker~$s_A \in S$, and the second by the speaker~$s_B \in S$.

\begin{align}
  &\text{Are the same people speaking in both} \notag \\
  &\text{recordings?} + d_A(s_{A_0}, s_{A_1}) + d_B(s_{B_0}, s_{B_1})
  \tag{CRSI\_2} \label{eq:crsi2}
\end{align}

\noindent
where

\begin{equation*}
  \begin{aligned}
    &s_{A_0}, s_{A_1}, s_{B_0}, s_{B_1} \in S \\
    &d_A, d_B \in D, \quad d_A \neq d_B \\
    &s_{A_0} \neq s_{A_1} \land s_{B_0} \neq s_{B_1}
  \end{aligned}
\end{equation*}

\noindent
The answer for~\ref{eq:crsi2} is \emph{Yes} if~$(s_{A_0}, s_{A_1}) = (s_{B_0}, s_{B_1}) \lor (s_{A_0}, s_{A_1}) = (s_{B_1}, s_{B_0})$.

Similarly, template~\ref{eq:crsi3} requires verifying if the same people are speaking in both dialogues, but in this case the order of speakers is also taken into account.

\begin{align}
  &\text{Do the people on both recordings speak in the} \notag \\
  &\text{same order?} + d_A(s_{A_0}, s_{A_1}) + d_B(s_{B_0}, s_{B_1})
  \tag{CRSI\_3} \label{eq:crsi3}
\end{align}

\noindent
where

\begin{equation*}
  \begin{aligned}
    &s_{A_0}, s_{A_1}, s_{B_0}, s_{B_1} \in S \\
    &d_A, d_B \in D, \quad d_A \neq d_B \\
    &s_{A_0} \neq s_{A_1} \land s_{B_0} \neq s_{B_1}
  \end{aligned}
\end{equation*}

\noindent
The answer for~\ref{eq:crsi3} is \emph{Yes} if~$s_{A_0} = s_{B_0} \land s_{A_1} = s_{B_1}$.

\subsection{Selective Text Inference}

Let~$G$ and~$S$ be the predefined sets of genders and speakers, respectively:

\begin{align*}
  G = \{ &\text{woman}, \text{man} \} \\
  S = \{ &s_1, s_2, s_3, s_4 \}
\end{align*}

\noindent
and~$U_0$ the predefined set of utterances:

\begin{itemize}
  \item ``He walked into the barber shop and requested a traditional shave and a haircut.''
  \item ``I saw her yesterday, she is pregnant.''
  \item ``This guy is playing computer games with his dog.''
  \item ``He adjusted his tie before heading to the office.''
  \item ``Person leaving the building was wearing a dress so she appeared to be a woman.''
  \item ``He tossed his keys onto the counter after walking into the house.''
\end{itemize}

\noindent
then:

\begin{itemize}
  \item $s(u)$ means that utterance~$u \in U_i, \quad i \in \{0, \dots, 3\}$ is spoken by speaker~$s \in S$
  \item $\mathrm{gender}(s) \in G$ is a function that returns the gender of the speaker~$s \in S$
  \item $\mathrm{talking\_about}(u, g)$ means that the utterance~$u \in U_i, \quad i \in \{0, \dots, 3\}$ is talking about gender~$g \in G$
\end{itemize}

Template~\ref{eq:sti0} requires determining whether the speaker~$s$ is talking about a person of the opposite gender, i.e. the answer is affirmative if~$\neg \mathrm{talking\_about} (u, \mathrm{gender}(s))$.

\begin{align}
  &\text{Is the person the speaker is talking about of} \notag \\
  &\text{the opposite gender?} + s(u)
  \tag{STI\_0} \label{eq:sti0}
\end{align}

\noindent
where~$s\in S$, $u\in U_0$.

Let~$B$ be the predefined set of subjects:

\begin{align*}
  B = \{ &\text{health}, \text{entertainment}, \text{environment}, \\
    &\text{politics}, \text{sports}, \text{technology} \}
\end{align*}

\noindent
and~$U_1$ the predefined set of utterances:

\begin{itemize}
  \item ``The constitution sets the rules for a country's government.''
  \item ``The Earth's environment is our responsibility to protect.''
  \item ``Theme parks and amusement parks offer thrilling rides and attractions.''
  \item ``Tennis players compete in Grand Slam tournaments.''
  \item ``Cloud computing has made it easier to store and access data remotely.''
  \item ``A good night's sleep is crucial for physical and mental recovery.''
  \item ``Conserving water is essential for our ecosystem.''
  \item ``Eating a balanced diet is essential for well-being.''
  \item ``Reading books is a great way to escape reality.''
\end{itemize}

\noindent
then~$\mathrm{subject}(u) \in B$ is a function that returns the subject of the utterance~$u \in U_1$.

Template~\ref{eq:sti1} involves assessing whether a speaker of the given gender is talking about the given subject. This requires the model to recognize gender of both speaker~$s_A$ and~$s_B$, identify if either of them is~$g \in G$, and if so, verify if they are talking about~$b \in B$.

\begin{align}
  &\text{Is the } g \text{ in the recording talking about } b \text{?} \notag \\
  &+ s_A(u_A) + s_B(u_B)
  \tag{STI\_1} \label{eq:sti1}
\end{align}

\noindent
where

\begin{equation*}
  \begin{aligned}
    &g \in G, \quad b \in B \\
    &s_A, s_B \in S, \quad s_A \neq s_B \\
    &u_A, u_B \in U_1, \quad u_A \neq u_B \\
    &\mathrm{subject}(u_A) \neq \mathrm{subject}(u_B) \\
    &\neg ( (\mathrm{gender}(s_A) = \mathrm{gender}(s_B) = g) \\
    &\quad \land (\mathrm{subject}(u_A) = \mathrm{subject}(u_B) = b) )
  \end{aligned}
\end{equation*}

\noindent
The answer to template~\ref{eq:sti1} is \emph{Yes} if

\begin{align*}
  &(\mathrm{gender}(s_A) = g \land \mathrm{subject}(u_A) = b) \\
  &\lor (\mathrm{gender}(s_B) = g \land \mathrm{subject}(u_B) = b)
\end{align*}

Let~$A$ be the predefined set of answers:

\begin{align*}
  A = \{ &\text{Ag}, \text{Alaska}, \text{Antarctica}, \text{blue}, \text{Denali}, \\
    &\text{Jupiter}, \text{ostrich}, \text{Saturn}, \text{Shakespeare}, \\
    &\text{Warsaw}, \text{whale} \}
\end{align*}

\noindent
and~$U_2$ the predefined set of questions:

\begin{itemize}
  \item ``What is the largest desert in the world?''
  \item ``What is the largest living species of bird?''
  \item ``What is the highest mountain peak in North America?''
  \item ``What is the largest mammal on Earth?''
  \item ``What planet has the most extensive rings?''
  \item ``What is the largest state in the United States?''
  \item ``What is the chemical symbol for silver?''
  \item ``Who wrote Romeo and Juliet?''
  \item ``What is the capital of Poland?''
  \item ``What is the largest planet in our solar system?''
  \item ``What color is the sky?''
\end{itemize}

\noindent
then~$\mathrm{answer}(u) \in A$ is a function that returns the answer to the question~$u \in U_2$.

In template~\ref{eq:sti2}, the model must verify if~$a \in A$ is an answer to the question asked by a speaker of the given gender. In this case, it must correctly recognize the gender of both speaker~$s_A$ and~$s_B$, and if either of them is~$g \in G$, answer their question and compare the response to~$a$.

\begin{align}
  &\text{Is } a \text{ a correct answer to the question asked} \notag \\
  &\text{by a } g \text{?} + s_A(u_A) + s_B(u_B)
  \tag{STI\_2} \label{eq:sti2}
\end{align}

\noindent
where

\begin{equation*}
  \begin{aligned}
    &s_A, s_B \in S, \quad s_A \neq s_B \\
    &u_A, u_B \in U_1, \quad u_A \neq u_B \\
    &\mathrm{subject}(u_A) \neq \mathrm{subject}(u_B) \\
    &\neg ((\mathrm{gender}(s_A) = \mathrm{gender}(s_B) = g) \\
    &\quad\land (\mathrm{answer}(u_A) = \mathrm{answer}(u_B) = a))
  \end{aligned}
\end{equation*}

\noindent
The answer is \emph{Yes} if

\begin{align*}
  (&\mathrm{gender}(s_A) = g \land \mathrm{answer}(u_A) = a) \\
  &\lor (\mathrm{gender}(s_B) = g \land \mathrm{answer}(u_B) = a)
\end{align*}

Let~$U_3$ be the predefined set of utterances:

\begin{itemize}
  \item ``Marie Curie was the first woman to win a Nobel Prize, awarded for her work in physics.''
  \item ``Ava enjoyed practicing yoga in the morning.''
  \item ``Leonardo da Vinci was a polymath and one of the most influential artists of the Renaissance.''
  \item ``Charles Darwin was a British naturalist who proposed the theory of evolution through natural selection.''
  \item ``Jackson was excited to go on his summer vacation.''
  \item ``Olivia loved to spend her free time reading books.''
  \item ``Benjamin Franklin was a leading figure in the American Enlightenment, known for his scientific and literary contribution.''
  \item ``Cinderella's fairy godmother helped her get ready for the ball.''
  \item ``Rapunzel's long hair was used as a ladder by a prince to reach her tower.''
  \item ``Rosa Parks was a civil rights activist who sparked the Montgomery Bus Boycott in 1955.''
\end{itemize}

\noindent
then~$\mathrm{historical}(u)$ means that utterance~$u \in U_3$ is about a famous historical figure.

Template~\ref{eq:sti3} involves verifying whether a speaker of the given gender is talking about a  famous historical figure. That means that the model must recognize the gender of both speakers~$s_A$ and~$s_B$, and if either of them is~$g \in G$, verify who is the person they are talking about.

\begin{align}
  &\text{Is the person the } g \text{ is talking about a famous} \notag \\
  &\text{historical figure?} + s_A(u_A) + s_B(u_B)
  \tag{STI\_3} \label{eq:sti3}
\end{align}

\noindent
where

\begin{equation*}
  \begin{aligned}
    &s_A, s_B\in S, \quad s_A \neq s_B \\
    &u_A, u_B\in U_3, \quad u_A \neq u_B \\
    &\mathrm{historical}(u_A) \Rightarrow \neg \mathrm{historical}(u_B)
  \end{aligned}
\end{equation*}

\noindent
In this case, the answer is \emph{Yes} if

\begin{align*}
  &(\mathrm{gender}(s_A) = g \land \mathrm{historical}(u_A)) \\
  &\lor (\mathrm{gender}(s_B) = g \land \mathrm{historical}(u_B))
\end{align*}

\subsection{Sound Reasoning}

Let~$S$ be the predefined set of sounds:

\begin{align*}
  S = \{ &\text{airplane}, \text{bell}, \text{bird}, \text{bird}_2, \text{cat}, \\
    &\text{coffee shop}, \text{concert}, \text{dog}, \text{elephant}, \\
    &\text{football game}, \text{horse}, \text{kids}, \text{motorcycle}, \\
    &\text{opera}, \text{rain}, \text{sea}, \text{sheep}, \text{thunderstorm}, \\
    &\text{traffic}, \text{train station} \}
\end{align*}

\noindent
then:

\begin{itemize}
  \item $\mathrm{animal}(s)$ is a function that implies that~$s\in S$ is a sound made by an animal
  \item $\mathrm{bigger}(a_1, a_2)$ is a function that implies that~$a_1$ is bigger than~$a_2$, where~$\mathrm{animal}(a_1) \land \mathrm{animal}(a_2)$
\end{itemize}

Template~\ref{eq:sr0} requires determining if the animal that makes the sound~$s$ is bigger or
smaller than the one mentioned in the question.

\begin{align}
  &\text{Is the animal that makes the following sound } \notag \\
  &z \text{ than } a \text{ ?} + s
  \tag{SR\_0} \label{eq:sr0}
\end{align}

\noindent
where

\begin{equation*}
  \begin{aligned}
    &a \in \{\text{bird}, \text{cat}, \text{dog}, \text{elephant}, \text{horse}, \text{sheep}\} \\
    &z \in \{\text{bigger}, \text{smaller}\} \\
    &s \in S, \quad \mathrm{animal}(s) \land s \neq a \\
    &a = \text{bird} \to \neg (s = \text{bird}_2) \\
    &a = \text{cat} \to \neg (s = \text{dog}) \\
    &a = \text{dog} \to \neg (s = \text{cat})
  \end{aligned}
\end{equation*}

\noindent
The answer to template~\ref{eq:sr0} is

\begin{align*}
  \text{Yes} \Leftrightarrow &(z = \text{bigger} \land \mathrm{bigger}(s, a)) \\
    &\lor (z = \text{smaller} \land \mathrm{bigger}(a, s))
\end{align*}

Template~\ref{eq:sr1} involves assessing whether the sounds heard indicate that the weather is nice.

\begin{align}
  &\text{Do the following sounds indicate that the} \notag \\
  &\text{weather is } d \text{ ?} + s
  \tag{SR\_1} \label{eq:sr1}
\end{align}

\noindent
where

\begin{equation*}
  \begin{aligned}
    &d \in \{\text{bad}, \text{nice}\} \\
    &s \in \{\text{bird}, \text{bird}_2, \text{rain}, \text{thunderstorm}\} \subseteq S
  \end{aligned}
\end{equation*}

Let~$\mathrm{is\_nice}(s)$ be a function that implies that sound~$s$ indicates that the weather is nice. Then, the answer to~\ref{eq:sr1} is affirmative if~$(d = \text{bad} \land \neg \mathrm{is\_nice}(s)) \lor (d = \text{nice} \land \mathrm{is\_nice}(s))$.

In template~\ref{eq:sr2}, the model is required to recognize an animal based on the given sound and determine whether one can ride such animal.

\begin{align}
  &\text{Can you ride an animal that makes this sound?} \notag \\
  &+ s
  \tag{SR\_2} \label{eq:sr2}
\end{align}

\noindent
where $s \in \{$bird, bird\textsubscript{2}, cat, dog, elephant, horse, sheep$\} \subseteq S$. Let~$\mathrm{is\_rideable}(s)$ be a function that implies that an animal making sound~$s$ is rideable. Then, the answer is \emph{Yes} if~$\mathrm{is\_rideable}(s)$.

Template~\ref{eq:sr3} involves assessing whether a thing making the given sound can fly.

\begin{align}
  &\text{Can a thing that makes the following sound fly?} \notag \\
  &+ s
  \tag{SR\_3} \label{eq:sr3}
\end{align}

\noindent
where~$s \in \{$airplane, bell, bird, bird\textsubscript{2}, cat, dog, elephant, horse, motorcycle, sheep$\} \subseteq S$. Let~$\mathrm{can\_fly}(s)$ be a function that implies that a thing making the sound~$s$ can fly. Then, the answer to~\ref{eq:sr3} is \emph{Yes} if~$\mathrm{can\_fly}(s)$.

Template~\ref{eq:sr4} is similar to ~\ref{eq:sr1}, as it requires the model to decide whether the sounds indicate that one should leave the house.

\begin{align}
  &\text{Should I leave the house now?} + s
  \tag{SR\_4} \label{eq:sr4}
\end{align}

\noindent
where~$s \in \{$bird, bird\textsubscript{2}, rain, thunderstorm$\} \subseteq S$. The answer is affirmative if~$\mathrm{is\_nice}(s)$.

For template~\ref{eq:sr5}, the model must determine whether the given sound can be heard at the given place. Let~$P$ be the predefined set of places:

\begin{align*}
  P = \{ &\text{at the airport}, \text{at the caf\'e}, \text{at the concert}, \\
    &\text{at the opera}, \text{at the school}, \text{at the stable}, \\
    &\text{at the stadium}, \text{at the train station}, \\
    &\text{at the zoo}, \text{by the sea}, \text{in the forest}, \\
    &\text{in the street} \}
\end{align*}

\noindent
then~$\mathrm{heard\_at}(s, p)$ implies that sound~$s \in S$ can be heard at place~$p \in P$.

\begin{align}
  &\text{Can the following sound be heard } p \text{ ?} + s
  \tag{SR\_5} \label{eq:sr5}
\end{align}

\noindent
where~$s \in \{$airplane, bird, coffee shop, concert, elephant, football game, horse, kids, opera, sea, traffic, train station$\} \subseteq S$.

Template~\ref{eq:sr6} requires determining if the vehicle that makes the given sound is bigger or smaller than the one mentioned in the question.

\begin{align}
  &\text{Is the vehicle making the following sound } \notag \\
  &z \text{ than } v \text{?} + s
  \tag{SR\_6} \label{eq:sr6}
\end{align}

\noindent
where

\begin{equation*}
  \begin{aligned}
    &z \in \{\text{bigger}, \text{smaller}\} \\
    &v \in \{\text{airplane}, \text{car}, \text{motorcycle}\} \\
    &s \in \{\text{airplane}, \text{motorcycle}\} \subseteq S, \quad s \neq v
  \end{aligned}
\end{equation*}

Let~$\mathrm{bigger}(v_1, v_2)$ be the function that implies that~$v_1$ is bigger than~$v_2$. Then, the answer to~\ref{eq:sr6} is \emph{Yes} if~$(z = \text{bigger} \land \mathrm{bigger}(s, v)) \lor (z = \text{smaller} \land \mathrm{bigger}(v, s))$.

In template~\ref{eq:sr7} the model must recognize an animal based on the sound and determine whether it is a mammal. Let~$\mathrm{class}(s) \in \{\text{aves}, \text{mammals}\}$ be a function that returns a biological class to which belongs the animal making the sound~$s \in \{ s' \in S \mid \mathrm{animal}(s) \}$.

\begin{align}
  &\text{Is the animal that makes the following sound} \notag \\
  &\text{a mammal?} + s
  \tag{SR\_7} \label{eq:sr7}
\end{align}

\noindent
where~$s \in \{ s' \in S \mid \mathrm{animal}(s') \}$. The answer is affirmative if~$\mathrm{class}(s) = \text{mammals}$.

Template~\ref{eq:sr8} requires similar reasoning, but there are two sounds to recognize.

\begin{align}
  &\text{Are both animals that make the following} \notag \\
  &\text{sounds mammals?} + s_A + s_B
  \tag{SR\_8} \label{eq:sr8}
\end{align}

\noindent
where~$s_A, s_B \in \{ s' \in S \mid \mathrm{animal}(s') \}$, $s_A \neq s_B$. Then, the answer is \emph{Yes} if~$\mathrm{class}(s_A) = \text{mammals} \land \mathrm{class}(s_B) = \text{mammals}$.

Similarly, template~\ref{eq:sr9} involves verifying whether two animals are from the same biological class based on the sounds they make.

\begin{align}
  &\text{Are both animals that make the following} \notag \\
  &\text{sounds from the same biological class?} + s_A \notag \\
  &+ s_B
  \tag{SR\_9} \label{eq:sr9}
\end{align}

\noindent
where~$s_A, s_B \in \{ s' \in S \mid \mathrm{animal}(s') \}$, $s_A \neq s_B$. The answer is \emph{Yes} if~$\mathrm{class}(s_A) = \mathrm{class}(s_B)$.

Template~\ref{eq:sr10} is requires recognition of two animals based on the sounds they make and determining whether both of them are bigger or smaller than the given animal.

\begin{align}
  &\text{Are both animals that make the following} \notag \\
  &\text{sounds } z \text{ than } a \text{?} + s_A + s_B
  \tag{SR\_10} \label{eq:sr10}
\end{align}

\noindent
where

\begin{equation*}
  \begin{aligned}
    &z \in \{\text{bigger}, \text{smaller}\} \\
    &a \in \{\text{bird}, \text{cat}, \text{dog}, \text{elephant}, \text{horse}, \text{sheep}\} \\
    &s_A, s_B \in \{ s' \in S \mid \mathrm{animal}(s') \} \\
    &s_A \neq s_B \neq a
  \end{aligned}
\end{equation*}

\noindent
The answer is \emph{Yes} if

\begin{align*}
  (&z = \text{bigger} \land \mathrm{bigger}(s_A, a) \land \mathrm{bigger}(s_B, a)) \\
  &\lor \\
  (&z = \text{smaller} \land \mathrm{bigger}(a, s_A) \land \mathrm{bigger}(a, s_B))
\end{align*}

In template~\ref{eq:sr11} the model must assess whether both animals making the given sounds are rideable.

\begin{align}
  &\text{Can you ride both animals that make these} \notag \\
  &\text{sounds?} + s_A + s_B
  \tag{SR\_11} \label{eq:sr11}
\end{align}

\noindent
where~$s_A, s_B \in \{ s' \in S \mid \mathrm{animal}(s') \}$, $s_A \neq s_B$. The answer is affirmative if~$\mathrm{is\_rideable}(s_A) \land \mathrm{is\_rideable}(s_B)$.

Similarly, template~\ref{eq:sr12} requires assessing if both things making the given sounds can fly.

\begin{align}
  &\text{Can both things that make the following} \notag \\
  &\text{sounds fly?} + s_A + s_B
  \tag{SR\_12} \label{eq:sr12}
\end{align}

\noindent
where~$s_A, s_B \in \{$airplane, bell, bird, bird\textsubscript{2}, cat, dog, elephant, horse, motorcycle, sheep$\} \subseteq S$, $s_A \neq s_B$. The answer to~\ref{eq:sr12} is \emph{Yes} if~$\mathrm{can\_fly}(s_A) \land \mathrm{can\_fly}(s_B)$.

Templates~\ref{eq:sr13} and~\ref{eq:sr14} involve verifying whether given sounds can be heard
indoors or outdoors.

\begin{align}
  &\text{Can this sound be usually heard } w \text{?} + s
  \tag{SR\_13} \label{eq:sr13} \\
  \notag \\
  &\text{Can both these sounds be usually heard } \notag \\
  &w \text{?} + s_A + s_B
  \tag{SR\_14} \label{eq:sr14}
\end{align}

\noindent
where

\begin{equation*}
  \begin{aligned}
    &w \in \{\text{indoors}, \text{outdoors}\} \\
    &s, s_A, s_B \in S, \quad s_A \neq s_B
  \end{aligned}
\end{equation*}

The answer to~\ref{eq:sr13} is affirmative if~$\mathrm{heard\_at}(s, w)$, and to~\ref{eq:sr14} if~$\mathrm{heard\_at}(s_A, w) \land \mathrm{heard\_at}(s_B, w)$.

\subsection{Speech Features Comparison}

Let~$S$, $C$, $G$, and~$A$ be the predefined sets of speakers, accents, genders, and ages, respectively:

\begin{align*}
  S = \{ &s_1, s_2, s_3, s_4, s_5, s_6, s_7, s_8, s_9, s_{10} \} \\
  C = \{ &\text{Australian}, \text{English}, \text{Scottish} \} \\
  G = \{ &\text{female}, \text{male} \} \\
  A = \{ &\text{fifties}, \text{twenties} \}
\end{align*}

\noindent
and~$U$ the predefined set of utterances:

\begin{itemize}
  \item ``The coffee smells amazing this morning.''
  \item ``I can't believe how bright the moon is tonight.''
  \item ``The birds chirped a gentle melody.''
\end{itemize}

\noindent
then:

\begin{itemize}
  \item $s(u)$ means that utterance~$u \in U$ is spoken by speaker~$s \in S$
  \item $\mathrm{accent}(s) \in C$ is a function that returns the accent of speaker~$s \in S$
  \item $\mathrm{gender}(s) \in G$ is a function that returns the gender of speaker~$s \in S$
  \item $\mathrm{age}(s) \in A$ is a function that returns the age of speaker~$s \in S$
\end{itemize}

Template~\ref{eq:sfc0} requires determining whether the second utterance is the same as the first one, but read with the given accent.

\begin{align}
  &\text{Is the second recording the same text but read} \notag \\
  &\text{with } c \text{ accent?} + s_A(u_A) + s_B(u_B)
  \tag{SFC\_0} \label{eq:sfc0}
\end{align}

\noindent
where

\begin{equation*}
  \begin{aligned}
    &c \in C \\
    &s_A, s_B \in S, \quad s_A \neq s_B \\
    &u_A, u_B \in U \\
    &\mathrm{accent}(s_A) = c \Rightarrow \neg (\mathrm{accent}(s_B) = c)
  \end{aligned}
\end{equation*}

The answer is affirmative if~$u_A = u_B \land \mathrm{accent}(s_B) = c$.

Template~\ref{eq:sfc1} involves verifying whether the second utterance is the same as the first one, but read with the given gender's voice.

\begin{align}
  &\text{Is the second recording the same text as the} \notag \\
  &\text{first recording but spoken by a } g \text{ voice?} \notag \\
  &+ s_A(u_A) + s_B(u_B)
  \tag{SFC\_1} \label{eq:sfc1}
\end{align}

\noindent
where

\begin{equation*}
  \begin{aligned}
    &g \in G \\
    &s_A, s_B \in S, \quad s_A \neq s_B \\
    &u_A, u_B \in U \\
    &\mathrm{gender}(s_A) = g \Rightarrow \neg (\mathrm{gender}(s_B) = g)
  \end{aligned}
\end{equation*}

The answer is \emph{Yes} if~$u_A = u_B \land \mathrm{gender}(s_B) = g$.

In template~\ref{eq:sfc2}, the model must recognize the gender of both speakers and verify if it is
the same.

\begin{align}
  &\text{Are both speakers the same gender?} + s_A(u_A) \notag \\
  &+ s_B(u_B)
  \tag{SFC\_2} \label{eq:sfc2}
\end{align}

\noindent
where~$s_A, s_B \in S$, $s_A \neq s_B$; $u_A, u_B \in U$, $u_A \neq u_B$. Then, the answer is \emph{Yes} if~$\mathrm{gender}(s_A) = \mathrm{gender}(s_B)$.

Similarly, template~\ref{eq:sfc3} requires recognition of the age of both speakers and verifying if
it the same.

\begin{align}
  &\text{Are both speakers the same age?} + s_A(u_A) \notag \\
  &+ s_B(u_B)
  \tag{SFC\_3} \label{eq:sfc3}
\end{align}

\noindent
where~$s_A, s_B \in S$, $s_A \neq s_B$; $u_A, u_B \in U$, $u_A \neq u_B$. Then, the answer is \emph{Yes} if~$\mathrm{age}(s_A) = \mathrm{age}(s_B)$.

\subsection{Text and Sound Reasoning}

Let~$S_P$, $S_A$, $S_O$ be the predefined sets of speakers, sounds of animals, and other sounds respectively:

\begin{align*}
  S_P = \{ &s_1, s_2, s_3, s_4 \} \\
  S_A = \{ &\text{bird}, \text{bird}_2, \text{cat}, \text{dog}, \text{elephant}, \text{horse}, \\
    &\text{sheep} \} \\
  S_O = \{ &\text{airplane}, \text{bell}, \text{kids}, \text{motorcycle}, \text{opera}, \\
    &\text{rain}, \text{thunderstorm}, \text{sea}, \text{concert}, \text{traffic} \} \\
\end{align*}

\noindent
$U_A$ the predefined set of utterances about animals:

\begin{itemize}
  \item ``The bird sings a cheerful melody at dawn.''
  \item ``The cat purs softly on the windowsill.''
  \item ``The dog wagged its tail happily.''
  \item ``The elephant drinks water with its long trunk.''
  \item ``A horse gallops across the open field.''
  \item ``The sheep grazed peacefully in the green meadow.''
\end{itemize}

\noindent
and~$U_O$ the predefined set of other utterances:

\begin{itemize}
  \item ``The airplane soars high above the clouds.''
  \item ``The bell rings, echoing through the hallway.''
  \item ``The kids laughed loudly as they ran around playing tag.''
  \item ``A motorcycle is a fast vehicle.''
  \item ``Opera singing tells dramatic stories.''
  \item ``Rain taps gently against the glass.''
  \item ``A thunderstorm rumbles loudly in the distance.''
  \item ``The waves crash against the shore.''
  \item ``A concert is a live performance of music.''
  \item ``The street was bustling with honking cars.''
\end{itemize}

\noindent
then:

\begin{itemize}
  \item $s_p(u)$ means that utterance~$u \in U_A \cup U_O$ is spoken by speaker~$s_p \in S$
  \item $\mathrm{about}(u)$ is a function that returns the main topic of the utterance~$u \in U_A \cup U_O$
\end{itemize}

Template~\ref{eq:tsr0} requires determining whether the spoken utterance is about the following sound.

\begin{align}
  &\text{Is the person talking about the following} \notag \\
  &\text{sound?} + s_p(u_o) + s_o
  \tag{TSR\_0} \label{eq:tsr0}
\end{align}

\noindent
where~$s_p \in S_P$, $u_o \in U_O$, and~$s_o \in S_O$. The answer is \emph{Yes} if~$\mathrm{about}(u_o) = s_o$.

Similarly, template~\ref{eq:tsr1} involves verifying whether the spoken utterance is about an animal making the following sound.

\begin{align}
  &\text{Is the person talking about an animal making} \notag \\
  &\text{the following sound?} + s_p(u_a) + s_A
  \tag{TSR\_1} \label{eq:tsr1}
\end{align}

\noindent
where~$s_p \in S_P$, $u_a \in U_A$, and~$s_a \in S_A$. The answer is \emph{Yes} if~$\mathrm{about}(u_a) = s_a$.

In template~\ref{eq:tsr2}, the model is required to recognize two animals based on an utterance and a sound, and then assess whether the second is smaller or bigger. Let~$\mathrm{bigger}(a_1, a_2)$ be a function that implies that animal~$a_1$ is bigger than~$a_2$.

\begin{align}
  &\text{Is the animal the person is talking about } z \notag \\
  &\text{than the one making the sound?} + s_p(u_a) \notag \\
  &+ s_a
  \tag{TSR\_2} \label{eq:tsr2}
\end{align}

\noindent
where

\begin{equation*}
  \begin{aligned}
    &z \in \{\text{bigger}, \text{smaller}\} \\
    &s_p \in S_P \\
    &u_a \in U_A, \quad s_a \in S_A, \quad \mathrm{about}(u_a) \neq s_a \\
    &s \neq \begin{cases}
      \text{bird}_2, & \text{if } \mathrm{about}(u_a) = \text{bird}, \\
      \text{bird}, & \text{if } \mathrm{about}(u_a) = \text{bird}_2, \\
      \text{dog}, & \text{if } \mathrm{about}(u_a) = \text{cat}, \\
      \text{cat}, & \text{if } \mathrm{about}(u_a) = \text{dog}.
    \end{cases}
  \end{aligned}
\end{equation*}

\noindent
The answer is \emph{Yes} if

\begin{align*}
  ( &z = \text{bigger} \land \mathrm{bigger}(\mathrm{about}(u_A), s_A) ) \\
  &\lor ( z = \text{smaller} \land \mathrm{bigger}(s_A, \mathrm{about}(u_A)) )
\end{align*}

Template~\ref{eq:tsr3} requires recognition of animals based on utterance and sound, and determining whether both of them are mammals. Let~$\mathrm{class}(a) \in \{\text{aves}, \text{mammals}\}$ be a function that returns a biological class to which an animal~$a$ belongs.

\begin{align}
  &\text{Are both the animal the person is talking about} \notag \\
  &\text{and the one making the sound mammals?} \notag \\
  &+ s_p(u_a) + s_a
  \tag{TSR\_3} \label{eq:tsr3}
\end{align}

\noindent
where

\begin{equation*}
  \begin{aligned}
    &s_p \in S_P \\
    &u_a \in U_A, \quad s_a \in S_A, \quad \mathrm{about}(u_a) \neq s_a \\
    &\mathrm{about}(u_a) = \text{bird} \Rightarrow s_a \neq \text{bird}_2
  \end{aligned}
\end{equation*}

\noindent
The answer is affirmative if~$\mathrm{class}(\mathrm{about}(u_a)) = \text{mammals} \land \mathrm{class}(s_a) = \text{mammals}$.

Similarly, template~\ref{eq:tsr4} involves determining whether both animals are from the same biological class.

\begin{align}
  &\text{Are both the animal the person is talking about} \notag \\
  &\text{and the one making the sound from the same} \notag \\
  &\text{biological class?} + s_p(u_a) + s_a
  \tag{TSR\_4} \label{eq:tsr4}
\end{align}

\noindent
where

\begin{equation*}
  \begin{aligned}
    &s_p \in S_P \\
    &u_a \in U_A, \quad s_a \in S_A, \quad \text{about}(u_a) \neq s_a \\
    &\mathrm{about}(u_a) = \text{bird} \Rightarrow s_a \neq \text{bird}_2
  \end{aligned}
\end{equation*}

\noindent
The answer is affirmative if~$\mathrm{class}(\mathrm{about}(u_a)) = \mathrm{class}(s_a)$.

In template~\ref{eq:tsr5}, the model must recognize animals based on utterance and sound, and verify if one can ride both of them. Let~$\mathrm{is\_rideable}(a)$ be a function that implies that animal~$a$ is rideable.

\begin{align}
  &\text{Can you ride both the animal the person is} \notag \\
  &\text{talking about and the one making the sound?} \notag \\
  &+ s_p(u_a) + s_a
  \tag{TSR\_5} \label{eq:tsr5}
\end{align}

\noindent
where

\begin{equation*}
  \begin{aligned}
    &s_p \in S_P \\
    &u_a \in U_A, \quad s_a \in S_A, \quad \mathrm{about}(u_a) \neq s_a \\
    &\mathrm{about}(u_a) = \text{bird} \Rightarrow s_a \neq \text{bird}_2
  \end{aligned}
\end{equation*}

\noindent
The answer is affirmative if~$\mathrm{is\_rideable}(\mathrm{about}(u_a)) \land \mathrm{is\_rideable}(s_a)$.

Template~\ref{eq:tsr6} involves recognition of two objects or animals, and determining whether both
of them can fly. Let~$\mathrm{can\_fly}(o)$ be a function that object or animal~$o$ can fly.

\begin{align}
  &\text{Can both the thing the person is talking about} \notag \\
  &\text{and the one making the sound fly?} + s_p(u) \notag \\
  &+ s
  \tag{TSR\_6} \label{eq:tsr6}
\end{align}

\noindent
where

\begin{equation*}
  \begin{aligned}
    &s_p \in S_P \\
    &u \in U' \subseteq U_A \cup U_O \\
    &\mathrm{about}(u) \neq s \\
    &\mathrm{about}(u) = \text{bird} \Rightarrow s \neq \text{bird}_2 \\
    &s \in \left\{
      \begin{aligned}
        &\text{bird}, \text{bird}_2, \text{cat}, \text{dog}, \\
        &\text{elephant}, \text{horse}, \text{sheep}, \\
        &\text{airplane}, \text{bell}, \text{motorcycle}
      \end{aligned}
    \right\} \subseteq S_A \cup S_O \\
  \end{aligned}
\end{equation*}

\noindent
and~$U'$ consists of:

\begin{itemize}
  \item ``The bird sings a cheerful melody at dawn.''
  \item ``The cat purs softly on the windowsill.''
  \item ``The dog wagged its tail happily.''
  \item ``The elephant drinks water with its long trunk.''
  \item ``A horse gallops across the open field.''
  \item ``The sheep grazed peacefully in the green meadow.''
  \item ``The airplane soars high above the clouds.''
  \item ``The bell rings, echoing through the hallway.''
  \item ``A motorcycle is a fast vehicle.''
\end{itemize}

\noindent
The answer is affirmative if~$\mathrm{can\_fly}(\mathrm{about}(u)) \land \mathrm{can\_fly}(s)$.

Template~\ref{eq:tsr7} requires assessing whether both objects, events or animals recognized based on utterance and sound, can be heard indoors or outdoors. Let~$\mathrm{heard\_at}(s, p)$ be a function that implies that object, event or animal~$s$ can be heard at place~$p$.

\begin{align}
  &\text{Can both the thing the person is talking about} \notag \\
  &\text{and the one making the sound be usually} \notag \\
  &\text{heard } p \text{ ?} + s_p(u) + s
  \tag{TSR\_7} \label{eq:tsr7}
\end{align}

\noindent
where

\begin{equation*}
  \begin{aligned}
    &p \in \{\text{indoors}, \text{outdoors}\} \\
    &s_p \in S_P \\
    &u \in U_A \cup U_O \\
    &s \in S_A \cup S_O \\
    &\mathrm{about}(u) \neq s \\
    &\mathrm{about}(u) = \text{bird} \Rightarrow s \neq \text{bird}_2
  \end{aligned}
\end{equation*}

\noindent
The answer to~\ref{eq:tsr7} is \emph{Yes} if~$\mathrm{heard\_at}(s, p) \land \mathrm{heard\_at}(\mathrm{about}(u), p)$.

\subsection{Text and Temporal Localization Reasoning}

Let~$Sp$, $P$, $Sn$ be the predefined sets of speakers, places, and sounds, respectively:

\begin{align*}
  Sp = \{ &s_1, s_2, s_3, s_4 \} \\
  P = \{ &\text{at the school}, \text{at the caf\'e}, \\
    &\text{at the train station}, \text{at the concert} \} \\
  Sn = \{ &\text{kids}, \text{coffee shop}, \text{train station}, \text{concert}, \\
    &\text{football game}, \text{traffic}, \text{sea}, \text{Christmas} \}
\end{align*}

\noindent
and~$D_0$ the predefined set of dialogues:

\begin{itemize}
  \item ``Did you understand what she meant by that last part?'' \\
    ``Not really. I'll need to check my notes later.''
  \item ``Do you want to sit by the windows or over here?'' \\
    ``Let's sit here. It's quieter.''
  \item ``Did you see that lighting effect during the last part?'' \\
    ``Yeah, it was incredible - it matched the rhythm perfectly!''
  \item ``How long do we have to wait now?'' \\
    ``Just a few more minutes, I think.''
\end{itemize}

\noindent
then:

\begin{itemize}
  \item $d(s_A, s_B)$ means that the first utterance in dialogue~$d \in D_0 \cup D_2 \cup D_3$ is spoken by the speaker~$s_A \in Sp$, and the second by the speaker~$s_B \in Sp$
  \item $\mathrm{place}(x) \in P$ is a function that returns a place where~$x \in Sn \cup D_0 \cup D_2 \cup D_3 \cup U$ can be heard
  \item $\frac{d(s_A, s_B)}{n}$ means that the background sound~$n \in Sn$ is added to the dialogue~$d(s_A, s_B)$; $d \in D_0 \cup D_2 \cup D_3$; $s_A, s_B \in Sp$
\end{itemize}

Template~\ref{eq:ttlr0} requires recognizing whether a conversation takes place in the given spot based on dialogue and background sounds.

\begin{align}
  &\text{Does the following conversation take place } \notag \\
  &p \text{ ?} + \frac{d(s_A, s_B)}{n}
  \tag{TTLR\_0} \label{eq:ttlr0}
\end{align}

\noindent
where

\begin{equation*}
  \begin{aligned}
    &p \in P \\
    &d \in D_0 \\
    &s_A, s_B \in Sp, \quad s_A \neq s_B \\
    &n \in \left\{
      \begin{aligned}
        &\text{kids}, \text{coffee shop}, \\
        &\text{concert}, \text{train station}
      \end{aligned}
    \right\} \subseteq Sn
  \end{aligned}
\end{equation*}

\noindent
The answer is affirmative if~$p = \mathrm{place}(d) = \mathrm{place}(n)$.

Let~$U$ be the predefined set of utterances:

\begin{itemize}
  \item ``The energy here is unreal, and the players are giving it their all!''
  \item ``This street is packed with honking cars and bustling crowds!''
  \item ``The waves crash against the shore, and the salty breeze is so refreshing!''
\end{itemize}

Template~\ref{eq:ttlr1} involves assessing whether a speaker is describing the acoustic scene that they are in.

\begin{align}
  &\text{Does the speaker describe the acoustic scene} \notag \\
  &\text{that they are in?} + \frac{s(u)}{n}
  \tag{TTLR\_1} \label{eq:ttlr1}
\end{align}

\noindent
where

\begin{equation*}
  \begin{aligned}
    &s \in Sp, \quad u \in U \\
    &n \in \{\text{football game}, \text{traffic}, \text{sea}\} \subseteq Sn
  \end{aligned}
\end{equation*}

\noindent
The answer is \emph{Yes} if~$\mathrm{place}(u) = \mathrm{place}(n)$.

Let~$D_2$ be the predefined set of dialogues for template~\ref{eq:ttlr2}:

\begin{itemize}
  \item ``Did you remember to pack sunscreen?'' \\
    ``Yeah, and I grabbed some snacks for the trip too.''
  \item ``Did you get everything on your list?'' \\
    ``Almost, but I still need to wrap a few things.''
\end{itemize}

\noindent
then~$\mathrm{time}(x) \in \{$Christmas, summer$\}$ is a function that returns a time when~$x \in Sn \cup D_2$ can be heard.

In template~\ref{eq:ttlr2}, the model is required to recognize the time of the year based on dialogue and background sound, and verify if it matches the given time.

\begin{align}
  &\text{Based on the sounds and conversation, is it } \notag \\
  &t \text{ now?} + \frac{d(s_A, s_B)}{n}
  \tag{TTLR\_2} \label{eq:ttlr2}
\end{align}

\noindent
where

\begin{equation*}
  \begin{aligned}
    &t \in \{\text{Christmas}, \text{summer}\} \\
    &d \in D_2 \\
    &s_A, s_B \in Sp, \quad s_A \neq s_B \\
    &n \in \{\text{Christmas}, \text{sea}\} \subseteq Sn
  \end{aligned}
\end{equation*}

\noindent
In this case, the answer is \emph{Yes} if~$\mathrm{time}(d) = \mathrm{time}(n) = t$.

Let~$D_3$ be the predefined set of dialogues for~\ref{eq:ttlr3}:

\begin{itemize}
  \item ``Hey, are you still coming over later?'' \\
    ``Yeah, I'll be there around six - traffic's a bit slow right now.''
  \item ``Do you want to grab something to eat later?'' \\
    ``Sure, how about that new place downtown?''
  \item ``Have you seen my phone anywhere?'' \\
    ``I think you left it on the kitchen counter.''
  \item ``Did you remember to water the plants today?'' \\
    ``Oh no, I completely forgot!''
  \item ``Did you hear about the new caf\'e downtown?'' \\
    ``No, what's it like?''
  \item ``Did you finish the group project?'' \\
    ``Almost, I just need to finalize the slides.''
\end{itemize}

Template~\ref{eq:ttlr3} involves verifying if the acoustic environment of two speakers is the same.

\begin{align}
  &\text{Is the acoustic environment of both people the} \notag \\
  &\text{same?} + d\left( \frac{s_A}{n_A}, \frac{s_B}{n_B} \right)
  \tag{TTLR\_3} \label{eq:ttlr3}
\end{align}

\noindent
where

\begin{equation*}
  \begin{aligned}
    &d \in D_3 \\
    &s_A, s_B \in Sp, \quad s_A \neq s_B \\
    &n_A, n_B \in Sn
  \end{aligned}
\end{equation*}

\noindent
For~\ref{eq:ttlr3}, the answer is affirmative if~$n_A = n_B$.

\section{Human Evaluation Results}
\label{app:human-evaluation-results}

Table~\ref{tab:human-evaluation} presents the accuracy per task achieved by three evaluators (\textbf{A}, \textbf{B}, and \textbf{C}) on the ART-H subset of the benchmark.

Each evaluator was presented with~216~prompts, to which they responded with a \emph{Yes} or \emph{No} answer. Overall, the tasks did not pose a challenge to any of the evaluators. The highest accuracy,~98.6\%, was achieved for \textbf{Audio Arithmetics}, \textbf{Selective Text Inference}, and \textbf{Text and Temporal Localization Reasoning}. Tasks requiring recognition of accents or languages (i.e., \textbf{Cross-Recording Language Identification} and \textbf{Speech Features Comparison}) resulted in the lowest accuracies:~80.6\% and~73.6\%, respectively.

All evaluators achieved a total accuracy of at least~90\%, resulting in a human baseline of~92.9\%.

\begin{table}
  \caption{Accuracy of the human evaluation of the ART benchmark.}
  \label{tab:human-evaluation}
  \centering
  \begin{tabular}{ccccc}
    \hline
    & \textbf{A} & \textbf{B} & \textbf{C} & \textbf{Average} \\
    \hline
    \textbf{AA} & 1.000 & 0.958 & 1.000 & 0.986 \\
    \textbf{ATD} & 1.000 & 0.958 & 0.917 & 0.958 \\
    \textbf{CRLI} & 0.958 & 0.625 & 0.833 & 0.806 \\
    \textbf{CRSI} & 1.000 & 0.917 & 1.000 & 0.972 \\
    \textbf{STI} & 1.000 & 0.958 & 1.000 & 0.986 \\
    \textbf{SR} & 1.000 & 0.958 & 0.958 & 0.972 \\
    \textbf{SFC} & 0.750 & 0.750 & 0.708 & 0.736 \\
    \textbf{TSR} & 0.958 & 1.000 & 0.917 & 0.958 \\
    \textbf{TTLR} & 1.000 & 1.000 & 0.958 & 0.986 \\
    \hline
    \textbf{Total} & 0.963 & 0.903 & 0.921 & 0.929 \\
    \hline
  \end{tabular}
\end{table}

\section{Experimental Setup}
\label{app:experimental-setup}

\begin{table*}[htbp]
  \caption{Inference parameters.}
  \label{tab:inference-params}
  \centering
  \begin{tabular}{llll}
    \hline
    \textbf{Model} & \textbf{Parameters} & & \\
    \hline
    \textbf{Whisper Large v3} & \makecell[l]{
      \texttt{max\_length}: 448 \\
      \texttt{return\_timestamps}: \texttt{false}
    } \\
    \hline
    \textbf{Llama 3.3} & \makecell[l]{
      \texttt{max\_model\_len}: 2048 \\
      \texttt{presence\_penalty}: 0.0 \\
      \texttt{frequency\_penalty}: 0.0
    } & \makecell[l]{
      \texttt{repetition\_penalty}: 1.0 \\
      \texttt{temperature}: 0.6 \\
      \texttt{top\_p}: 0.9
    } & \makecell[l]{
      \texttt{top\_k}: 0 \\
      \texttt{min\_p}: 0.0
    } \\
    \hline
    \textbf{Qwen3} & \makecell[l]{
      \texttt{enable\_thinking}: \texttt{false} \\
      \texttt{max\_model\_len}: 2048 \\
      \texttt{presence\_penalty}: 0.0
    } & \makecell[l]{
      \texttt{frequency\_penalty}: 0.0 \\
      \texttt{repetition\_penalty}: 1.0 \\
      \texttt{temperature}: 0.6
    } & \makecell[l]{
      \texttt{top\_p}: 0.95 \\
      \texttt{top\_k}: 20 \\
      \texttt{min\_p}: 0.0
    } \\
    \hline
    \textbf{Audio Flamingo 3} & \makecell[l]{
      \texttt{max\_length}: 20 \\
      \texttt{length\_penalty}: 1.0 \\
    } & \makecell[l]{
      \texttt{repetition\_penalty}: 1.0 \\
      \texttt{temperature}: 1.0
    } & \makecell[l]{
      \texttt{top\_k}: 50 \\
      \texttt{top\_p}: 1.0
    } \\
    \hline
    \textbf{GAMA} & \makecell[l]{
      \texttt{max\_new\_tokens}: 400 \\
      \texttt{do\_sample}: \texttt{true}
    } & \makecell[l]{
      \texttt{temperature}: 0.1 \\
      \texttt{top\_p}: 0.95
    } & \makecell[l]{
      \texttt{top\_k}: 500
    } \\
    \hline
    \textbf{Qwen-Audio-Chat} & \makecell[l]{
      \texttt{max\_new\_tokens}: 512 \\
      \texttt{do\_sample}: \texttt{true}
    } & \makecell[l]{
      \texttt{top\_p}: 0.5 \\
      \texttt{top\_k}: 0
    }\\
    \hline
    \textbf{Qwen2-Audio} & \makecell[l]{
      \texttt{max\_model\_len}: 2048 \\
      \texttt{presence\_penalty}: 0.0 \\
      \texttt{frequency\_penalty}: 0.0
    } & \makecell[l]{
      \texttt{repetition\_penalty}: 1.0 \\
      \texttt{temperature}: 0.7 \\
      \texttt{top\_p}: 0.5
    } & \makecell[l]{
      \texttt{top\_k}: 20 \\
      \texttt{min\_p}: 0.0
    } \\
    \hline
    \textbf{Ultravox v0.4.1} & \makecell[l]{
      \texttt{max\_model\_len}: 2048 \\
      \texttt{presence\_penalty}: 0.0 \\
      \texttt{frequency\_penalty}: 0.0 \\
    } & \makecell[l]{
      \texttt{repetition\_penalty}: 1.0 \\
      \texttt{temperature}: 1.0 \\
      \texttt{top\_p}: 1.0
    } & \makecell[l]{
      \texttt{top\_k}: 0 \\
      \texttt{min\_p}: 0.0
    } \\
    \hline
    \textbf{Ultravox v0.6} & \makecell[l]{
      \texttt{max\_model\_len}: 2048 \\
      \texttt{presence\_penalty}: 0.0 \\
      \texttt{frequency\_penalty}: 0.0 \\
    } & \makecell[l]{
      \texttt{repetition\_penalty}: 1.0 \\
      \texttt{temperature}: 1.0 \\
      \texttt{top\_p}: 1.0
    } & \makecell[l]{
      \texttt{top\_k}: 0 \\
      \texttt{min\_p}: 0.0
    } \\
    \hline
  \end{tabular}
\end{table*}

\begin{figure*}[t]
\centering
\begin{lstlisting}[
  language=,caption={System prompt used for the LLM-as-a-judge approach.},
  label={lst:judge-prompt},basicstyle=\small\ttfamily,breaklines=true,frame=single
]
  You will receive a question, an expected answer (which can always be reduced to "Yes" or "No"), and an actual answer that was given. Your task is to check wheter the given answer is relevant to the question. If it is relevant, respond with "Relevant" on the first line, followed by another line stating whether the given answer is correct or incorrect compared to the expected answer. If it is irrelevant, respond with "Irrelevant" on the first line, followed by an explanation on the next line.
  Example Input:
  Question: Are there as many dog barks as there are bell rings?
  Expected Answer: No
  Given Answer: There are more dog barks than bell rings.
  Example Output:
  Relevant
  Incorrect
  Example Input:
  Question: Is one of the recordings a quieter version of the other one?
  Expected Answer: Yes
  Given Answer: Yes, the second recording is a shortened version of the first one.
  Example Output:
  Irrelevant
  The answer does not address the question about the quieter audio.
\end{lstlisting}
\end{figure*}

Table~\ref{tab:inference-params} presents the parameters used during inference. Each model was prompted using the default settings suggested by the authors.

The system prompt used for the evaluation with LLM as a judge is shown in Listing~\ref{lst:judge-prompt}. The same prompt was used for both Llama 3.3 and Qwen3.

\begin{lstlisting}[
  language=,caption={General text prompt used for the inference in \emph{Yes/No} experiments.},
  label={lst:yn-prompt},basicstyle=\small\ttfamily,breaklines=true,frame=single,breakindent=0pt
]
  Answer the question from the audio. Answer only "Yes" or "No".
\end{lstlisting}

\begin{lstlisting}[
  language=,caption={General text prompt used for the inference in \emph{Descriptive} experiments.},
  label={lst:descriptive-prompt},basicstyle=\small\ttfamily,breaklines=true,frame=single,breakindent=0pt
]
  Answer the question from the audio.
\end{lstlisting}

The general text prompt used for the experiments with \emph{Yes/No} approach is shown in Listing~\ref{lst:yn-prompt}, and with \emph{Descriptive} approach in Listing~\ref{lst:descriptive-prompt}

The GAMA model required modification of the general prompt to improve the quality of the generated results. The prompts used in this case are shown in Listings~\ref{lst:gama-yn-prompt} and~\ref{lst:gama-descriptive-prompt}.

\begin{lstlisting}[
  language=,caption={Text prompt used for the inference on GAMA model in \emph{Yes/No} experiments.},
  label={lst:gama-yn-prompt},basicstyle=\small\ttfamily,breaklines=true,frame=single,breakindent=0pt
  ]
  You will receive an audio sample containing a spoken question. Your task is to provide a concise and accurate answer. Answer only "Yes" or "No".
\end{lstlisting}

\begin{lstlisting}[
  language=,caption={Text prompt used for the inference on GAMA model in \emph{Descriptive} experiments.},
  label={lst:gama-descriptive-prompt},basicstyle=\small\ttfamily,breaklines=true,frame=single,breakindent=0pt
]
  You will receive an audio sample containing a spoken question. Your task is to provide a concise and accurate answer. Answer only the question contained in the audio without adding unnecessary details.
\end{lstlisting}

\section{Analysis of the results of experiments with the \emph{Descriptive} approach}
\label{app:descriptive-results}

\begin{table*}[htbp]
  \caption{Absolute accuracy per task on the ART benchmark using \emph{Descriptive} approach and Llama 3.3 as a judge.}
  \label{tab:descriptive-results-per-task-llama}
  \centering
  \begin{tabular}{m{0.2\textwidth}ccccccccc}
    \hline
    \textbf{Model} & \textbf{AA} & \textbf{ATD} & \textbf{CRLI} & \textbf{CRSI} & \textbf{STI} &
      \textbf{SR} & \textbf{SFC} & \textbf{TSR} & \textbf{TTLR} \\
    \hline
    Whisper + Llama & 0.158 & 0.157 & 0.126 & 0.275 & \textbf{0.503} & 0.240 & 0.603 & 0.591 & \textbf{0.522} \\
    Whisper + Qwen & \textbf{0.451} & \textbf{0.390} & \textbf{0.268} & \textbf{0.394} & 0.446 & \textbf{0.609} & \textbf{0.671} & \textbf{0.646} & 0.516 \\
    \hline
    Audio Flamingo 3 & 0.046 & 0.015 & 0.042 & 0.026 & 0.031 & 0.061 & 0.053 & 0.034 & 0.043 \\
    GAMA & 0.000 & 0.000 & 0.000 & 0.000 & 0.000 & 0.020 & 0.006 & 0.051 & 0.003 \\
    Qwen-Audio-Chat & 0.052 & 0.011 & 0.054 & 0.000 & 0.003  & 0.136 & 0.261 & 0.149 & 0.034 \\
    Qwen2-Audio\newline \scriptsize{(zero-shot)} & 0.439 & 0.382 & 0.496 & 0.385 & \textbf{0.501} & 0.453 & 0.481 & 0.460 & 0.443 \\
    Qwen2-Audio\newline \scriptsize{(one-shot, same template)} & 0.425 & \textbf{0.397} & \textbf{0.513} & \textbf{0.397} & 0.480 & \textbf{0.560} & 0.492 & 0.525 & \textbf{0.472} \\
    Qwen2-Audio\newline \scriptsize{(one-shot, different template)} & \textbf{0.446} & 0.301 & 0.494 & 0.374 & 0.482 & 0.466 & 0.500 & \textbf{0.559} & 0.426 \\
    Ultravox v0.4.1 & 0.158 & 0.022 & 0.319 & 0.220 & 0.281 & 0.126 & 0.436 & 0.346 & 0.250 \\
    Ultravox v0.6 & 0.405 & 0.352 & 0.286 & 0.176 & 0.372 & 0.394 & \textbf{0.566} & 0.540 & 0.462 \\
    \hline
    \textbf{Average} & 0.281 & 0.224 & 0.284 & 0.247 & 0.341 & 0.334 & \textbf{0.446} & \textbf{0.430} & 0.347 \\
    \hline
  \end{tabular}
\end{table*}

\begin{table*}[htbp]
  \caption{Absolute accuracy per task on the ART benchmark using \emph{Descriptive} approach and Qwen3 as a judge.}
  \label{tab:descriptive-results-per-task-qwen}
  \centering
  \begin{tabular}{m{0.2\textwidth}ccccccccc}
    \hline
    \textbf{Model} & \textbf{AA} & \textbf{ATD} & \textbf{CRLI} & \textbf{CRSI} & \textbf{STI} &
      \textbf{SR} & \textbf{SFC} & \textbf{TSR} & \textbf{TTLR} \\
    \hline
    Whisper + Llama & 0.125 & 0.008 & 0.128 & 0.491 & \textbf{0.559} & 0.233 & \textbf{0.639} & 0.724 & 0.676 \\
    Whisper + Qwen & \textbf{0.660} & \textbf{0.512} & \textbf{0.280} & \textbf{0.513} & 0.466 & \textbf{0.645} & 0.626 & \textbf{0.728} & \textbf{0.697} \\
    \hline
    Audio Flamingo 3 & 0.054 & 0.018 & 0.042 & 0.026 & 0.033 & 0.057 & 0.049 & 0.032 & 0.042 \\
    GAMA & 0.000 & 0.000 & 0.000 & 0.000 & 0.000 & 0.004 & 0.087 & 0.045 & 0.003 \\
    Qwen-Audio-Chat & 0.065 & 0.028 & 0.037 & 0.000 & 0.010 & 0.088 & 0.112 & 0.125 & 0.028 \\
    Qwen2-Audio\newline \scriptsize{(zero-shot)} & 0.542 & \textbf{0.353} & \textbf{0.604} & 0.373 & \textbf{0.502} & 0.492 & 0.490 & 0.411 & 0.448 \\
    Qwen2-Audio\newline \scriptsize{(one-shot, same template)} & 0.541 & 0.344 & 0.522 & 0.426 & 0.474 & \textbf{0.579} & 0.527 & 0.517 & 0.566 \\
    Qwen2-Audio\newline \scriptsize{(one-shot, different template)} & \textbf{0.554} & 0.307 & 0.544 & \textbf{0.463} & 0.491 & 0.506 & 0.477 & 0.532 & 0.477 \\
    Ultravox v0.4.1 & 0.100 & 0.003 & 0.197 & 0.186 & 0.205 & 0.064 & 0.290 & 0.228 & 0.250 \\
    Ultravox v0.6 & 0.521 & 0.340 & 0.260 & 0.207 & 0.427 & 0.399 & \textbf{0.587} & \textbf{0.615} & \textbf{0.626} \\
    \hline
    \textbf{Average} & 0.345 & 0.211 & 0.286 & 0.296 & 0.348 & 0.334 & \textbf{0.426} & \textbf{0.436} & 0.419 \\
    \hline
  \end{tabular}
\end{table*}

Table~\ref{tab:descriptive-results-per-task-llama} shows the accuracy of each model on each of the nine tasks as evaluated by Llama 3.3 as a judge.

Overall, the \textbf{Speech Features Comparison} and \textbf{Text and Sound Reasoning} tasks have the highest average accuracy scores, suggesting that these tasks are easier for the models. In contrast, \textbf{Audio Transformation Detection}, \textbf{Cross-Recording Speaker Identification}, and \textbf{Audio Arithmetics} have the lowest average accuracy scores, indicating greater ambiguity or difficulty.

Performance varies considerably across tasks at the model level. Several models demonstrate task-specific strengths, achieving high accuracy on certain tasks but performing poorly on others. This indicates limited generalization across task types.

Table~\ref{tab:descriptive-results-per-task-qwen} shows the accuracy obtained by each model as evaluated by Qwen3 as a judge.

Once again \textbf{Speech Features Comparison} and \textbf{Text and Sound Reasoning} have the highest average accuracies, with \textbf{Text and Temporal Localization Reasoning} close behind. In contrast, \textbf{Audio Transformation Detection} and \textbf{Cross-Recording Language Identification} have the lowest average accuracy scores.

Overall, evaluation with Qwen3 as a judge reinforces the presence of systematic differences in task difficulty. These results largely mirror the trends observed with Llama 3.3 as a judge, while also highlighting differences in absolute accuracy levels across tasks that depend on the method of evaluation.

\begin{table*}[htbp]
  \caption{Judgment alignment between human, Llama 3.3, and Qwen3.}
  \label{tab:human-llm-alignment}
  \centering
  \begin{tabular}{lccc}
    \hline
    \textbf{Model} & \textbf{Human-Llama 3.3} & \textbf{Human-Qwen3} & \textbf{Llama 3.3-Qwen3} \\
    \hline
    Whisper + Llama & 65.28\% & 60.65\% & 73.15\% \\
    Whisper + Qwen & 72.22\% & 61.11\% & 79.63\% \\
    \hline
    Audio Flamingo 3 & 78.70\% & 75.00\% & 75.93\% \\
    GAMA & 99.54\% & 98.15\% & 98.61\% \\
    Qwen-Audio-Chat & 87.50\% & 80.09\% & 85.19\% \\
    Qwen2-Audio \scriptsize{(zero-shot)} & 69.44\% & 62.04\% & 82.87\% \\
    Qwen2-Audio \scriptsize{(one-shot, same template)} & 80.56\% & 71.76\% & 84.72\%\\
    Qwen2-Audio \scriptsize{(one-shot, different template)} & 72.22\% & 64.35\% & 80.09\% \\
    Ultravox v0.4.1 & 47.22\% & 54.17\% & 68.06\% \\
    Ultravox v0.6 & 72.69\% & 61.57\% & 78.70\% \\
    \hline
  \end{tabular}
\end{table*}

\section{Human-LLM Alignment}
\label{app:human-llm-alignment}

The level of agreement among a human judge and two LLM-based judges was evaluated on the ART-H subset of the benchmark. All of the systems were evaluated using the \emph{Descriptive} approach. Table~\ref{tab:human-llm-alignment} shows the results, which measure agreement as the percentage of aligned judge decisions.

Overall, the agreement between the two LLM judges was consistently higher than the agreement between either LLM judge and the human evaluator. The Llama~3.3-Qwen3 alignment ranged from~68.06\% to~98.61\%, whereas the Human-Llama 3.3 agreement ranged from~47.22\% to~99.54\%, and the Human-Qwen3 agreement ranged from~54.17\% to~98.15\%. These results suggest that the two LLM judges have more similar evaluation criteria than the human judge.

There were significant model-level differences. The GAMA model exhibited near-perfect agreement across all judge pairings~($\geq 98\%$), indicating strong consistency and minimal ambiguity in evaluation outcomes. However, it should be noted that the human evaluator deemed all of this model's responses irrelevant. In contrast, the Ultravox v0.4.1 model showed substantially lower agreement with the human judge (47.22\% with Llama~3.3 and~54.17\% with Qwen3), despite moderate agreement between the two LLM judges (68.06\%). This difference stems from the fact that this model tended to provide speculative answers based on its knowledge of the real world.

Qwen2-Audio, Ultravox v0.6 and both cascaded systems exhibited intermediate agreement levels, where the alignment between LLM judges remained relatively high~(73.15-84.72\%), while human-LLM agreement was more variable. Notably, Audio Flamingo 3 and Qwen-Audio-Chat achieved comparatively strong human-LLM alignment~($\geq~75\%$), suggesting closer alignment between automated and human evaluation in these cases.

In summary, the results suggest that LLM judges are consistent within themselves but not aligned with human judgments. While some models demonstrate strong agreement, others reveal substantial divergence. This underscores the importance of model-specific analysis and careful interpretation of LLM-based evaluation outcomes.

\section{Error Analysis}
\label{app:error-analysis}

\subsection{\emph{Yes/No} approach}

Table~\ref{tab:error-analysis-yes-no} presents an error analysis for the \emph{Yes/No} approach based on human evaluation conducted on the ART-H subset of the benchmark.

The most common errors across models were related to failures in understanding or processing the audio input. Several models often failed to recognize speech or sound altogether. For example, GAMA model exhibites this behavior in~47.83\% of erroneous cases, while Ultravox v0.4.1 exhibited this behavior in an even higher percentage of cases, at~66.67\%. In Qwen2-Audio with zero-shot approach, failure to recognize speech or sound accounted for half of the observed errors. These results suggest limitations in audio perception or interpretation for these models.

Another common error was providing responses that were unrelated to the task. These responses constituted~50\% of errors for Qwen2-Audio with zero-shot approach, 27.59\% with one-shot approach and example from the same template, 24.05\% with one shot-approach and example from a different template, and~33.33\% for Ultravox v0.4.1 model. This pattern suggests that when the models failed to interpret the input properly, they often relied on unconstrained generation rather than explicitly signaling uncertainty.

Some models exhibited systematic, yet task-inappropriate, behaviors. For example, Qwen-Audio-Chat consistently returned a transcription of the audio instead of answering the question, accounting for~100\% of its errors. This indicates a significant bias toward speech-to-text functionality that potentially overrides the intended question-answering objective. Similarly, Qwen2-Audio with one-shot approach, frequently produced responses identifying the speaker (44.83\% and~35.44\%), indicating confusion between speaker recognition and the intended task.

\begin{table}[htbp]
  \caption{Error analysis of the model responses for \emph{Yes/No} approach based on human
    evaluation.}
  \label{tab:error-analysis-yes-no}
  \centering
  \begin{tabular}{lc}
    \hline
    \textbf{GAMA} \\
    \hspace{0.4em}Did not recognize the question & 51.30\% \\
    \hspace{0.4em}Did not recognize speech or sound & 47.83\% \\
    \hspace{0.4em}Random & 0.87\% \\
    \hline
    \textbf{Qwen-Audio-Chat} \\
    \hspace{0.4em}Transcription & 100.00\% \\
    \hline
    \textbf{Qwen2-Audio} \scriptsize{(zero-shot)} \\
    \hspace{0.4em}Did not recognize speech or sound & 50.00\% \\
    \hspace{0.4em}Random & 50.00\% \\
    \hline
    \textbf{Qwen2-Audio} \scriptsize{(one-shot, same template)} \\
    \hspace{0.4em}Speaker recognition & 44.83\% \\
    \hspace{0.4em}Random & 27.59\% \\
    \hspace{0.4em}Responded in different language & 10.34\% \\
    \hspace{0.4em}Did not recognize speech or sound & 10.34\% \\
    \hspace{0.4em}Cannot help & 6.90\% \\
    \hline
    \textbf{Qwen2-Audio} \scriptsize{(one-shot, different template)} \\
    \hspace{0.4em}Speaker recognition & 35.44\% \\
    \hspace{0.4em}Random & 24.05\% \\
    \hspace{0.4em}Responded in different language & 21.52\% \\
    \hspace{0.4em}Did not recognize speech or sound & 16.46\% \\
    \hspace{0.4em}Returned timestamps & 2.53\% \\
    \hline
    \textbf{Ultravox v0.4.1} \\
    \hspace{0.4em}Did not recognize speech or sound & 66.67\% \\
    \hspace{0.4em}Random & 33.33\% \\
    \hline
  \end{tabular}
\end{table}

Language-related errors were also observed. Qwen2-Audio with one-shot approach responded in a different language~10.34\% and~21.52\% of the time. This highlights inconsistencies in language control under failure conditions. Other notable errors were less frequent but still significant, including an explicit refusal or inability to help (6.90\% for the same template) and returning timestamps instead of answers (2.53\% for a different template).

Overall, the error analysis reveals that failures in \emph{Yes/No} approach are primarily due to misinterpretation or non-recognition of the audio input and systematic task confusion, in which the models default to transcription, speaker identification, or an unrelated generation.

Audio Flamingo 3, Ultravox v0.6, and the cascaded systems were excluded from this analysis, as they produced nearly~100\% relevant answers across all 9\,000 samples of the ART benchmark.

\subsection{\emph{Descriptive} approach}

Tables~\ref{tab:error-analysis-descriptive-cascaded} and~\ref{tab:error-analysis-descriptive-audiollms} provide an error analysis for the \emph{Descriptive} approach, which is based on human evaluation conducted on the ART-H subset of the benchmark.

\begin{table}[htbp]
  \caption{Error analysis of the cascaded systems' responses for \emph{Descriptive} approach based on human
    evaluation.}
  \label{tab:error-analysis-descriptive-cascaded}
  \centering
  \begin{tabular}{lc}
    \hline
    \textbf{Whisper + Llama} \\
    \hspace{0.4em}Cannot help & 44.65\% \\
    \hspace{0.4em}Asked for transcription & 30.19\% \\
    \hspace{0.4em}Did not recognize the question & 15.09\% \\
    \hspace{0.4em}Responded in different language & 9.43\% \\
    \hspace{0.4em}Speculative answer & 0.63\% \\
    \hline
    \textbf{Whisper + Qwen} \\
    \hspace{0.4em}Did not recognize the question & 49.32\% \\
    \hspace{0.4em}Cannot help & 30.14\% \\
    \hspace{0.4em}Responded in different language & 20.55\% \\
    \hline
  \end{tabular}
\end{table}

A common error for the \emph{Descriptive} approach was failing to recognize the intent of the question. This behavior was particularly common among Whisper + Qwen~(49.32\%) and Ultravox~v0.6~(45.36\%). In many cases, this lack of recognition resulted in fallback behaviors rather than an outright refusal to complete the task.

\begin{table}[htbp]
  \caption{Error analysis of the AudioLLMs' responses for \emph{Descriptive} approach based on human
    evaluation.}
  \label{tab:error-analysis-descriptive-audiollms}
  \centering
  \begin{tabular}{lc}
    \hline
    \textbf{Audio Flamingo 3} \\
    \hspace{0.4em}Transcription & 60.82\% \\
    \hspace{0.4em}Random & 21.65\% \\
    \hspace{0.4em}Did not recognize the question & 17.53\% \\
    \hline
    \textbf{GAMA} \\
    \hspace{0.4em}Random & 71.76 \\
    \hspace{0.4em}Did not recognize the question & 28.24 \\
    \hline
    \textbf{Qwen-Audio-Chat} \\
    \hspace{0.4em}Transcription & 78.65\% \\
    \hspace{0.4em}Did not recognize the question & 12.35\% \\
    \hspace{0.4em}Random & 8.99\% \\
    \hline
    \textbf{Qwen2-Audio} \scriptsize{(zero-shot)} \\
    \hspace{0.4em}Responded in different language & 61.11\% \\
    \hspace{0.4em}Random & 17.13\% \\
    \hspace{0.4em}Speaker recognition & 10.65\% \\
    \hspace{0.4em}Political matters & 6.48\% \\
    \hspace{0.4em}Cannot help & 2.78\% \\
    \hspace{0.4em}Returned timestamps & 1.39\% \\
    \hspace{0.4em}Did not recognize speech or sound & 0.46\% \\
    \hline
    \textbf{Qwen2-Audio} \scriptsize{(one-shot, same template)} \\
    \hspace{0.4em}Speaker recognition & 41.63\% \\
    \hspace{0.4em}Random & 35.41\% \\
    \hspace{0.4em}Responded in different language & 8.13\% \\
    \hspace{0.4em}Did not recognize speech or sound & 7.18\% \\
    \hspace{0.4em}Cannot help & 4.31\% \\
    \hspace{0.4em}Returned timestamps & 3.35\% \\
    \hline
    \textbf{Qwen2-Audio} \scriptsize{(one-shot, different template)} \\
    \hspace{0.4em}Random & 37.74\% \\
    \hspace{0.4em}Speaker recognition & 36.32\% \\
    \hspace{0.4em}Responded in different language & 10.38\% \\
    \hspace{0.4em}Did not recognize speech or sound & 7.08\% \\
    \hspace{0.4em}Returned timestamps & 4.72\% \\
    \hspace{0.4em}Cannot help & 3.77\% \\
    \hline
    \textbf{Ultravox v0.4.1} \\
    \hspace{0.4em}Speculative answer & 48.19\% \\
    \hspace{0.4em}Described the prompt & 30.57\% \\
    \hspace{0.4em}Respondend in different language & 10.88\% \\
    \hspace{0.4em}Cannot help & 9.33\% \\
    \hspace{0.4em}Random & 1.04\% \\
    \hline
    \textbf{Ultravox v0.6} \\
    \hspace{0.4em}Did not recognize the question & 45.36\% \\
    \hspace{0.4em}Cannot help & 35.05\% \\
    \hspace{0.4em}Speculative answer & 13.40\% \\
    \hspace{0.4em}Transcription & 3.09\% \\
    \hspace{0.4em}Responded in different language & 2.06\% \\
    \hspace{0.4em}Random & 1.03\% \\
    \hline
  \end{tabular}
\end{table}

\begin{table*}[htbp]
  \caption{Results of the random sampling analysis using \emph{Yes/No} approach.}
  \label{tab:sampling-yn}
  \centering
  \begin{tabular}{m{0.20\textwidth}|ccc|ccc}
    \hline
    & \multicolumn{3}{c|}{\textbf{Relevant}} & \multicolumn{3}{c}{\textbf{Absolute Accuracy}} \\
    \textbf{Model} & \textbf{mean} & \textbf{min} & \textbf{max} & \textbf{mean} & \textbf{min}
      & \textbf{max} \\
    \hline
    Whisper + Llama & 99.92 $\pm$ 0.19 & 98.89 & 100.00 & 0.5410 $\pm$ 0.0307 & 0.4704 & 0.6185 \\
    Whisper + Qwen & 99.92 $\pm$ 0.10 & 99.63 & 100.00 & 0.5619 $\pm$ 0.0248 & 0.5102 & 0.6194 \\
    \hline
    Audio Flamingo 3 & 100.00 $\pm$ 0.00 & 100.00 & 100.00 & 0.5446 $\pm$ 0.0301 & 0.4537
      & 0.6204 \\
    GAMA & 42.26 $\pm$ 1.62 & 38.15 & 46.30 & 0.2144 $\pm$ 0.0137 & 0.1843 & 0.2546 \\
    Qwen-Audio-Chat & 64.03 $\pm$ 1.54 & 60.19 & 67.22 & 0.3315 $\pm$ 0.0213 & 0.2750 & 0.3898 \\
    Qwen2-Audio\newline \scriptsize{(zero-shot)} & 85.06 $\pm$ 1.92 & 80.28 & 90.93
      & 0.4419 $\pm$ 0.0239 & 0.3694 & 0.4972 \\
    Qwen2-Audio\newline \scriptsize{(one-shot, same template)} & 87.27 $\pm$ 1.56 & 83.70 & 90.74
      & 0.4704 $\pm$ 0.0263 & 0.3991 & 0.5241 \\
    Qwen2-Audio\newline \scriptsize{(one-shot, different template)} & 67.93 $\pm$ 2.26 & 63.33
      & 72.59 & 0.3485 $\pm$ 0.0229 & 0.2880 & 0.4000 \\
    Ultravox v0.4.1 & 89.17 $\pm$ 1.19 & 85.93 & 92.31 & 0.4707 $\pm$ 0.0207 & 0.4278 & 0.5194 \\
    Ultravox v0.6 & 99.82 $\pm$ 0.25 & 98.98 & 100.00 & 0.5315 $\pm$ 0.0237 & 0.4722 & 0.5935 \\
    \hline
  \end{tabular}
\end{table*}

Another common error was models providing transcriptions instead of answering the question. This was particularly evident with Audio Flamingo 3 and Qwen-Audio-Chat, which returned transcriptions~60.82\% and~78.65\% of the time, respectively. This suggests a persistent bias toward speech-to-text behavior.

Several models exhibited a substantial proportion of random or incoherent responses, most notably GAMA~(71.76\%) and Qwen2-Audio with one-shot approach~(35.41\% and~37.74\%). These responses suggest that models often generate unconstrained outputs when uncertain rather than signaling uncertainty or failure.

Qwen2-Audio often confused the given task with speaker recognition. Language control failures were observed across multiple models, including Qwen2-Audio, Whisper + Qwen, and Ultravox v0.4.1. Notable but less common errors included speculative answers given by both Ultravox models, answering the prompt with a description instead of a direct response, and providing an unrelated safety-style refusal. For example, Qwen2-Audio with zero-shot approach stated that it could not discuss political manners, despite the task being unrelated.

Overall, the observed error patterns for the \emph{Descriptive} approach suggest that failures are primarily caused by a misunderstanding of the task objective, strong transcription and speaker identification biases, and an increased rate of speculative or random generation. Compared to the \emph{Yes/No} approach, these findings indicate that the \emph{Descriptive} approach includes a broader diversity of error categories, highlighting the need for improved task grounding.

\section{ART-H Robustness}
\label{app:sampling}

Since ART benchmark consists of~9\,000~samples, which equals to over~30~hours of audio, we designed ART-H. It is a subset of 24 samples per task, where half of the instances has \emph{Yes} answers and the rest has \emph{No} answers.

To ensure that the models' performance is stable across samples, a random sampling analysis was conducted. For each approach, 216 outputs were selected randomly 100 times, computing the number of relevant answers and the accuracy for each sample. The results are shown in Tables~\ref{tab:sampling-yn}--\ref{tab:sampling-descriptive-qwen}.

The resulting accuracies closely match the values achieved on the full set of~9\,000~outputs, demonstrating that the models' performance is consistent even on randomly selected subsets of outputs.

\begin{table*}[htbp]
  \caption{
    Results of the random sampling analysis using \emph{Descriptive} approach and Llama 3.3 as a
    judge.
  }
  \label{tab:sampling-descriptive-llama}
  \centering
  \begin{tabular}{m{0.20\textwidth}|ccc|ccc}
    \hline
    & \multicolumn{3}{c|}{\textbf{Relevant}} & \multicolumn{3}{c}{\textbf{Absolute Accuracy}} \\
    \textbf{Model} & \textbf{mean} & \textbf{min} & \textbf{max} & \textbf{mean} & \textbf{min}
      & \textbf{max} \\
    \hline
    Whisper + Llama & 62.35 $\pm$ 2.37 & 56.02 & 67.76 & 0.3529 $\pm$ 0.0265 & 0.2824 & 0.4120 \\
    Whisper + Qwen & 89.69 $\pm$ 1.58 & 85.45 & 93.43 & 0.4871 $\pm$ 0.0321 & 0.4000 & 0.5748 \\
    \hline
    Audio Flamingo 3 & 74.21 $\pm$ 2.92 & 65.26 & 80.93 & 0.3562 $\pm$ 0.0286 & 0.2817 & 0.4507 \\
    GAMA & 1.28 $\pm$ 0.74 & 0.00 & 3.70 & 0.0084 $\pm$ 0.0061 & 0.0000 & 0.0278 \\
    Qwen-Audio-Chat & 14.51 $\pm$ 1.94 & 8.80 & 18.06 & 0.0766 $\pm$ 0.0160 & 0.0370 & 0.1157 \\
    Qwen2-Audio\newline \scriptsize{(zero-shot)} & 92.35 $\pm$ 1.59 & 88.32 & 95.77
      & 0.4476 $\pm$ 0.0267 & 0.3632 & 0.5209 \\
    Qwen2-Audio\newline \scriptsize{(one-shot, same template)} & 94.01 $\pm$ 1.79 & 89.10 & 97.64
      & 0.4720 $\pm$ 0.0249 & 0.4140 & 0.5592 \\
    Qwen2-Audio\newline \scriptsize{(one-shot, different template)} & 93.18 $\pm$ 1.69 & 87.68
      & 97.66 & 0.4463 $\pm$ 0.0283 & 0.3814 & 0.5093 \\
    Ultravox v0.4.1 & 60.15 $\pm$ 3.17 & 52.09 & 68.84 & 0.2403 $\pm$ 0.0210 & 0.1831 & 0.3023 \\
    Ultravox v0.6 & 79.41 $\pm$ 2.40 & 73.49 & 85.51 & 0.3965 $\pm$ 0.0299 & 0.3224 & 0.4744 \\
    \hline
  \end{tabular}
\end{table*}

\begin{table*}[htbp]
  \caption{
    Results of the random sampling analysis using \emph{Descriptive} approach and Qwen3 as a judge.
  }
  \label{tab:sampling-descriptive-qwen}
  \centering
  \begin{tabular}{m{0.20\textwidth}|ccc|ccc}
    \hline
    & \multicolumn{3}{c|}{\textbf{Relevant}} & \multicolumn{3}{c}{\textbf{Absolute Accuracy}} \\
    \textbf{Model} & \textbf{mean} & \textbf{min} & \textbf{max} & \textbf{mean} & \textbf{min}
      & \textbf{max} \\
    \hline
    Whisper + Llama & 62.25 $\pm$ 2.06 & 56.94 & 67.59 & 0.3974 $\pm$ 0.0284 & 0.3380 & 0.4630 \\
    Whisper + Qwen & 87.90 $\pm$ 1.63 & 83.80 & 91.67 & 0.5691 $\pm$ 0.0320 & 0.4907 & 0.6435 \\
    \hline
    Audio Flamingo 3 & 77.10 $\pm$ 2.36 & 70.37 & 81.94 & 0.3555 $\pm$ 0.0269 & 0.2870 & 0.4398 \\
    GAMA & 2.20 $\pm$ 0.90 & 0.46 & 5.09 & 0.0133 $\pm$ 0.0071 & 0.0000 & 0.0324 \\
    Qwen-Audio-Chat & 19.92 $\pm$ 2.45 & 12.04 & 26.85 & 0.0529 $\pm$ 0.0153 & 0.0139 & 0.0972 \\
    Qwen2-Audio\newline \scriptsize{(zero-shot)} & 94.95 $\pm$ 1.28 & 92.59 & 98.15
      & 0.4667 $\pm$ 0.0263 & 0.4120 & 0.5370 \\
    Qwen2-Audio\newline \scriptsize{(one-shot, same template)} & 96.43 $\pm$ 1.36 & 91.67 & 99.07
      & 0.4993 $\pm$ 0.0257 & 0.4583 & 0.5787 \\
    Qwen2-Audio\newline \scriptsize{(one-shot, different template)} & 95.32 $\pm$ 1.28 & 91.67
      & 98.15 & 0.4819 $\pm$ 0.0278 & 0.4167 & 0.5417 \\
    Ultravox v0.4.1 & 50.80 $\pm$ 3.00 & 43.98 & 57.41 & 0.1707 $\pm$ 0.0226 & 0.1250 & 0.2222 \\
    Ultravox v0.6 & 77.55 $\pm$ 2.73 & 71.76 & 83.80 & 0.4449 $\pm$ 0.0298 & 0.3657 & 0.5231 \\
    \hline
  \end{tabular}
\end{table*}

\end{document}